\documentclass[notitlepage,11pt,amsmath,preprintnumbers,superscriptaddress,nofootinbib,aps] 
{revtex4-1}

\usepackage{amsmath}
\usepackage{lineno} 
\usepackage{ulem}

\usepackage{todonotes}

\presetkeys{todonotes}{color=green!40, size=\footnotesize}{}
\usepackage{hyperref}

\hypersetup{colorlinks=true,citecolor=blue,citebordercolor=red,linktoc=all,linkcolor=blue}

\usepackage{setspace}

\usepackage{amssymb,amsfonts,multirow}

\usepackage{hyperref}
\usepackage{multirow}
\usepackage{graphicx}
\usepackage{amsmath}
\usepackage{amssymb}
\usepackage{color,bm}

\long\def\del #1 \enddel { }

\usepackage{xcolor}
\usepackage{colortbl}

\definecolor{Gray}{gray}{0.85}
\definecolor{LightGreen}{rgb}{0.88, 1, 0.88}
\definecolor{Blue}{rgb}{0,1,1}
\definecolor{Lime}{rgb}{0,1,0}
\definecolor{LightCyan}{rgb}{0.88,1,1}
\definecolor{LightRed}{rgb}{1, 0.85, 0.85}
\definecolor{Red}{rgb}{1, 0, 0}
\definecolor{LightYellow}{rgb}{1, 1, 0.85}
\definecolor{Yellow}{rgb}{1,1,0.05}
\definecolor{LightBlue}{rgb}{0.87, 0.94, 1}
\definecolor{white}{gray}{1}
\definecolor{black}{gray}{0}

\definecolor{LightGray}{gray}{0.93}

\newcolumntype{G}{>{\columncolor{LightGray}}c}
\newcolumntype{L}{>{\columncolor{LightGray}}l}

\usepackage{graphicx}
\usepackage{amsmath}
\usepackage{amssymb}

\usepackage{epsfig}

\def\beq{\begin{equation}}
\def\eeq{\end{equation}}
\def\bea{\arraycolsep .1em \begin{eqnarray}}
\def\eea{\end{eqnarray}}
\def\tr{{\rm Tr}}

\def\arcsinh{{\text{arcsinh}}}

\def\half{{\frac{1}{2}}}

\def\eq#1{(\ref{#1})}

\def\s0#1#2{\mbox{\small{$ \frac{#1}{#2} $}}}
\def\0#1#2{\frac{#1}{#2}}

\usepackage{array,mathtools,amssymb,booktabs}
\newcolumntype{C}{>{$}c<{$}}
\AtBeginDocument{
\heavyrulewidth=.16em
\lightrulewidth=.1em
\cmidrulewidth=.03em
\belowrulesep=.4ex
\belowbottomsep=0pt
\aboverulesep=.4ex
\abovetopsep=0pt
\cmidrulesep=\doublerulesep
\cmidrulekern=.5em
\defaultaddspace=.5em
}
\makeatletter
    \def\CT@@do@color{%
      \global\let\CT@do@color\relax
            \@tempdima\wd\z@
            \advance\@tempdima\@tempdimb
            \advance\@tempdima\@tempdimc
    \advance\@tempdimb\tabcolsep
    \advance\@tempdimc\tabcolsep
    \advance\@tempdima2\tabcolsep
            \kern-\@tempdimb
            \leaders\vrule
                    \hskip\@tempdima\@plus  1fill
            \kern-\@tempdimc
            \hskip-\wd\z@ \@plus -1fill }

   \makeatother
   
\usepackage{setspace}

\begin{document}

\title{Asymptotic safety of scalar field theories}
\author{Daniel F.~Litim}
\address{
Department of Physics and Astronomy, University of Sussex, BN1 9QH, Brighton, UK}
\author{Matthew J.~Trott}
\address{
Department of Physics and Astronomy, University of Sussex, BN1 9QH, Brighton, UK}
\address{
\mbox{
SUPA, School of Physics and Astronomy, University of St Andrews, KY16 9SS, St Andrews, UK}}

\begin{abstract}
We  study $3d$
$O(N)$  symmetric 
scalar field theories   
using Polchinski's   renormalisation group. 
 In the infinite $N$ limit the model is solved exactly including at strong coupling.
 At short distances the theory is described by a line of 
 asymptotically safe
 ultraviolet
 fixed points
 bounded by asymptotic freedom at weak,   and the Bardeen-Moshe-Bander 
 phenomenon at strong sextic coupling.
 The Wilson-Fisher fixed point arises as an isolated low-energy fixed point. 
Further results  include the  conformal window 
for asymptotic safety,  convergence-limiting poles in the complex field plane, and the phase diagram with regions of first and second order phase transitions. 
We  substantiate a duality between
Polchinski's and Wetterich's versions of the functional renormalisation group,  also showing that that eigenperturbations are identical at any fixed point. At  a critical sextic coupling, the duality is worked out in detail to explain the spontaneous breaking of scale symmetry responsible for the generation of a light dilaton. Implications  for
asymptotic safety in other theories are indicated. 
\end{abstract}

\pagestyle{plain} \setcounter{page}{1}

\maketitle

\begin{spacing}{.81}
\tableofcontents
\end{spacing}

\section{\bf Introduction}
Fixed points of the renormalisation group play a fundamental role in quantum field theory and statistical physics \cite{Wilson:1971dc}. Infrared fixed points  control the long distance and low momentum behaviour of theories and are often associated with continuous phase transitions \cite {ZinnJustin:1996cy}. Ultraviolet (UV) fixed points  serve as a definition of quantum field theory, such as in asymptotic freedom \cite{Gross:1973id,Politzer:1973fx} or in asymptotic safety 
\cite{Weinberg:1980gg,Litim:2011cp,Litim:2014uca,Bond:2016dvk}. Interacting UV fixed points also ensure that the renormalisation group evolution of couplings remains finite even at shortest distances, opening new opportunities to define fundamental theories  including quantum gravity \cite{Weinberg:1980gg,Litim:2011cp}.

Much progress has been made in the understanding of asymptotic safety  
for  particle theories involving gauge fields, fermions, and scalars. 
At weak coupling,  necessary and sufficient conditions  for asymptotic safety   \cite{Bond:2016dvk,Bond:2017sem} and  strict no-go theorems \cite{Bond:2018oco} are known, with examples including simple \cite{Litim:2014uca,Buyukbese:2017ehm,Bond:2017tbw}, semi-simple \cite{Bond:2017wut,Bond:2017lnq,Kowalska:2017fzw}, and supersymmetric  gauge theories \cite{Bond:2017suy}. 
It has also been conjectured  that asymptotic safety may exist at strong coupling \cite{PalanquesMestre:1983zy,Gracey:1996he,Holdom:2010qs},  
In $3d$, the interest in  fixed points of relativistic quantum fields is motivated by  thermal and quantum phase transitions of spin liquids and quantum magnets, superconductors, topological insulators, and models for new materials such as graphene. These  can often be described by scalar, fermionic, and Yukawa models, and gauged or supersymmetric versions thereof \cite{Pelissetto:2000ek,Gies:2009hq,Braun:2010tt,
Litim:2011bf,Heilmann:2012yf,
Jakovac:2013jua,Jakovac:2014lqa,
Vacca:2015nta,Safari:2017tgs,
Ihrig:2018ojl,Lee:2018udi,Gracey:2018qba,Katsis:2018bvc}.
Some of the recent interest in $3d$ models is further fuelled by  a conjecture for novel types of dualities \cite{Aharony:2013dha,Aharony:2018pjn}, much in the spirit of Seiberg duality, or particle-vortex duality \cite{Peskin:1977kp} between  $O(2)$  magnets and  the abelian-Higgs model for superconductivity  \cite{Bergerhoff:1995zq,Bergerhoff:1995zm}. 

In this paper, we  study interacting  fixed points  and asymptotic safety of $3d$ $O(N)$ symmetric field theories in the large-$N$ limit. There are several reasons for doing so. Firstly,  in the limit of infinite $N$, the model is exactly solvable which makes it an ideal testing ground for the phenomenon of asymptotic safety in a purely scalar quantum field theory \cite{David:1985zz,Litim:1995ex,Tetradis:1995br,Litim:2016hlb,Marchais:2017jqc}. 
Secondly, the model is known to display  a variety of  low- and high-energy fixed points,  first and second order phase transitions, and  the spontaneous breaking of (scale) symmetry.
Previous work has covered aspects of symmetry breaking and $1/N$ corrections  \cite{Coleman:1974jh,Townsend:1975kh,
Litim:2016hlb,Juttner:2017cpr,Marchais:2017jqc},  stability of the ground state  and UV fixed points  
\cite{Appelquist:1981sf,Appelquist:1982vd,Pisarski:1982vz,Bardeen:1983rv,Litim:2016hlb,Juttner:2017cpr,Marchais:2017jqc}, and extensions with supersymmetry or Chern-Simons interactions \cite{Bardeen:1984dx,Eyal:1996da,Litim:2011bf,Heilmann:2012yf,Aharony:2012ns,Bardeen:2014paa}. 

Methodologies thus far have included
 perturbation theory, variational methods \cite{Bardeen:1983rv,Bardeen:1983st,David:1984we,David:1985zz,Amit:1984ri,Ananos:1996kk}, functional renormalisation \cite{Berges:2000ew,Comellas:1997tf,Litim:2016hlb,Juttner:2017cpr,Marchais:2017jqc}, and the conformal bootstrap \cite{Poland:2018epd}. 
The main technical novelty of this work is the   use of Polchinski's  renormalisation group, originally employed for  proofs of   renormalisability  \cite{POLCHINSKI1984269}. It is based on a Wilsonian UV cutoff, and allows a complete  understanding of conformal fixed points, convergence-limiting poles, universality of eigenperturbations, and the full  phase diagram including at strong coupling. Findings are then compared with earlier works \cite{Comellas:1997tf,Marchais:2017jqc}, in particular with results from   Wetterich's  functional renormalisation group \cite{Wetterich:1992yh} based on an IR momentum cutoff. A   duality between Polchinski's and Wetterich's equation \cite{Litim:2005us,Morris:2005ck} is exploited to explain the spontaneous breaking of scale invariance and the origin of mass. We also highlight qualitative links with  asymptotically safe $4d$ models  in  large-$N$ limits.

The outline of the paper is as follows. In Sec.~\ref{RG}, we recall Polchinski's and Wetterich's renormalisation group, their duality, and the basic equations of this paper. The beta functions for all couplings are solved locally and globally in Sec.~\ref{RC}. 
Sec.~\ref{IR} investigates infrared fixed points and convergence-limiting singularities in the complex field plane. Asymptotically safe ultraviolet fixed points and strong coupling effects are studied in
Sec.~\ref{UV}. 
The spontaneous breaking of scale symmetry and the Polchinski-Wetterich duality are analysed in Sec.~\ref{BMB}.
Sec.~\ref{PD} investigates the universal eigenperturbations, and the phase diagram including the conformal window with asymptotic safety and regions with first order phase transitions.
Sec.~\ref{Discussion} contains a summary of results and implications for asymptotic safety of particle theories in four dimensions.

\section{\bf Renormalisation group}\label{RG}

In this section we introduce two closely related exact functional renormalisation group equations, the Polchinski flow equation and the  Wetterich flow equation. We also provide the relevant expressions in the local potential approximation and explain an exact  duality between both setups.

\subsection{Polchinski equation}
The Polchinski flow equation \cite{POLCHINSKI1984269} for a Wilsonian action $S_k[\varphi_a]$ is an exact, functional equation based on an UV momentum cutoff $k$. It has initially been introduced   to facilitate proofs of perturbative renormalisability \cite{POLCHINSKI1984269,Ball:1993zy} solely using Wilson's renormalisation group \cite{Wilson:1973jj}.  It is given by
\begin{widetext}
\begin{equation}\label{floweqn}
\partial_t S_k[\varphi_a]=\frac{1}{2}\tr\;\partial_t P_k(q)\left[S^{(1)}_k(q)S^{(1)}_k(-q)-S^{(2)}_k(q,-q)-2(P_k(q))^{-1}\varphi(q)S^{(1)}_k(q)\right]
\end{equation}
\end{widetext}
where the $n$-point functions are defined as $S_k^{(n)}(q)=\delta^n S_k[\varphi_a]/\delta^n\varphi_a(q)$, and 
\beq\label{PkPol}
P_k(q)=\0{K(q^2/k^2)}{q^2}
\eeq 
denotes the massless cut-off propagator.\footnote{In the Polchinski RG literature, the RG momentum scale $k$ is often denoted as $\Lambda$. In this paper, we use $\Lambda$ to denote a reference energy scale, typically a ``high scale'' in the UV.}
 Also, $t=\ln(k/\Lambda)$ with $\Lambda$ is a suitable 
 reference energy scale, and the trace relates to a momentum integration. The function $K(q^2/k^2)$ suppresses high momentum modes with $K(q^2/k^2\rightarrow\infty)\rightarrow0$, and allows low momentum modes to propagate freely through $K(q^2/k^2\rightarrow0)\rightarrow1$. As such, the Wilsonian momentum scale $k$ takes the role of an UV cutoff. The exact scale dependence \eq{floweqn} of the interaction functional $S_k[\varphi_a]$ is strictly induced by the scale dependence of $K(q^2/k^2)$ \cite{doi:10.1142/S0217751X94000972}. We also note that Polchinski's equation \eq{floweqn} is related to Wilson's original flow \cite{Wilson:1973jj} by an appropriate rescaling.

The Polchinski flow has been used in many studies previously 
(e.g.~\cite{
Nicoll:1974zza,
Ball:1993zy,
Ball:1994ji,
Morris:1994ie,
Comellas1997539,
Bagnuls:2000ae,
Litim:2001dt,
Nandori:2002ri,
Kubyshin:2002zx,
Litim:2005us,
Bervillier:2005za,
ODwyer:2007brp,
Osborn:2009vs,
Osborn201229} and references therein).
  Here, we examine an $O(N)$ symmetric scalar field $\varphi_a$ in $d$ euclidean dimensions, the $(\phi^2)^3_{d}$ theory for short. The model has 
   previously been examined via the Polchinski equation within the derivative expansion \cite{Comellas1997539,Bagnuls:2000ae,Kubyshin:2002zx,Osborn:2009vs,Bervillier:2013hha}, and
using Wetterich's renormalisation group  in the optimised form of Litim \cite{Litim:2001up,Litim:2002cf,Litim:2005us,Litim:2010tt}. 
The model becomes exactly solvable in the infinite $N$ limit  \cite{Litim:1995ex,Tetradis:1995br,TETRADIS1994541,Litim:2005us,Marchais:2017jqc}. We will formulate an exact solution to the Polchinski flow at infinite $N$ in three-dimensions and apply some of the techniques developed earlier to inspect fixed points in the Legendre formalism, presenting our analysis in analogy to \cite{Litim:2005us,Litim:2016hlb,Juttner:2017cpr,Marchais:2017jqc}, and to aid comparison of Polchinski and Wetterich flow equations.

We work in the local potential approximation (LPA), considering only a uniform field configuration and do not renormalise the kinetic term \cite{Nicoll:1974zza}, which is well-justified in the large-$N$ limit. The Wilsonian interaction action becomes $S_k=\int_xV_k(\varphi_a)$,
and the Polchinski flow equation (\ref{floweqn}) reduces to 
\begin{equation}\label{dtV}
\partial_t V_k=\frac{1}{k^2}\left[\gamma\left(\frac{\partial V_k}{\partial\varphi_a}\right)^2-\alpha\frac{\partial^2 V_k}{\partial\varphi_a^2}\right]\,.
\end{equation}
The coefficients 
\beq
\label{scheme}
\gamma=-K'(0)\,,\quad \alpha=-\int \frac{d^dq}{(2\pi)^d}\, K'(q^2/k^2)
\eeq 
in \eq{dtV} are non-universal and depend on the choice of the Wilsonian regulator function $K$ \cite{Ball:1994ji,Litim:2005us}. Monotonically decreasing cutoff functions  $K(q^2/k^2)$ 
imply non-vanishing and positive coefficients \eq{scheme}.  Vanishing or negative RG coefficients $\alpha$ and  $\gamma$ may arise  for non-monotonically decreasing  (e.g.~oscillating) regulator functions $K$. For the sake of this work, and also to avoid regulator-induced artefacts
  \cite{Osborn201229}, we  limit ourselves to RG schemes with positive coefficients $\alpha$ and $\gamma$.\footnote{The possibility that  interacting fixed points slip into an unphysical regime due to the non-monotonicity of cutoff functions has previously been observed  in large-$N$ studies \cite{Eyal:1996da,Moshe:2003xn}, and in the context of lattice regularisations \cite{Kessler:1985ge}.} 
A particularly useful "optimised" choice for $K$ is given by
\beq\label{Kopt}
K_{\rm opt}\left(y\right)=\left(1-y\right)\theta\left(1-y\right)
\eeq
where $y={q^2}/{k^2}$. Evidently, the scale $k$ acts as a UV cutoff with $K_{\rm opt}\equiv 0$ for $y>1$ and $K_{\rm opt}=1-y$ for $y<1$. The choice \eq{Kopt} leads to the positive parameters
\beq\label{gaopt}
\gamma_{\rm opt}=1\,,\quad \alpha_{\rm opt}=\frac{2}{d\,L_d}\,k^d
\eeq
with $L_d=(4\pi)^{d/2}\Gamma(d/2)$  the $d$-dimensional loop factor \cite{Litim:2016hlb}. 
Provided that $\alpha$ and $\gamma$ in \eq{scheme} are positive numbers such as in \eq{gaopt}, we rescale  the potential $V_k\rightarrow \alpha v/\gamma $ and field $\varphi_a\rightarrow \sqrt{\alpha/k^d}\varphi_a$, and introduce the dimensionless variables $v(\tilde\varphi)=k^{-d}V_k(\varphi)$ and  $\tilde\varphi_a=k^{1-d/2}\varphi_a$. This leads to the manifestly scheme-independent Polchinski flow
\begin{equation}\label{vflow}
\partial_t v=-dv+\half(d-2)\tilde\varphi_a \frac{\partial v}{\partial \tilde\varphi_a}+\left(\frac{\partial v}{\partial \tilde\varphi_a}\right)^2- \frac{\partial^2 v}{\partial \tilde\varphi_a^2}.
\end{equation}
It is convenient to introduce the variable  $\rho=\half\tilde\varphi_a\tilde\varphi_a$ which is invariant under reflection in field space. The symmetry of our $O(N)$ model allows us to make the choice
$\tilde\varphi_a=\tilde\varphi \,\delta_{1a}$, 
meaning that $\tilde\varphi_a$ points into the 1-direction in field space.
We then make the transformation to a $\rho$ dependent potential $u(\rho)=v(\tilde\varphi_a)$ such that the second derivative term performing the trace becomes
\begin{equation}
\frac{\partial^2 v}{\partial \tilde\varphi_a^2}=\delta_{aa}u'+\tilde\varphi_a\tilde\varphi_a u''=Nu'+2\rho u'' .
\end{equation}
Primes denote derivatives with respect to $\rho$. Making the change of variables in (\ref{vflow}) we find
\begin{equation}\label{incneqn}
\partial_t u=-d u + (d-2)\rho u' +2\rho(u')^2-N u'-2\rho u'' .
\end{equation}
We then rescale the terms $\rho\rightarrow\rho/N$ and $v\rightarrow v/N$ and see that the second derivative term vanishes for $N\rightarrow\infty$. Finally we fix our theory to three dimensions to obtain
\begin{equation}\label{highneqn}
\partial_t u=-3 u + (\rho-1) u' +2\rho(u')^2 \,.
\end{equation}
Taking the derivative of (\ref{highneqn}) with respect to $\rho$ we obtain 
\begin{equation}\label{polsca}
\partial_t u'=-2u'+(\rho-1)u''+2(u')^2+4\rho u'u'' .
\end{equation}
This is the central equation of our study. Below, we investigate its global fixed point solutions for all fields and all couplings.

\subsection{Wetterich equation}\label{ERG}

The Wetterich  equation is an exact, functional flow equation based on a IR momentum cutoff  for the effective average  action $\Gamma_k$. It has first  been given in  \cite{Wetterich:1992yh,Ellwanger:1993mw,Morris:1993qb}  and reads
\begin{equation}\label{FRG}
\partial_t\Gamma_k=\frac12\tr\frac{1}{\Gamma_k^{(2)}+R_k}\partial_t R_k\,.
\end{equation}
It expresses the change of  $\Gamma_k$ with renormalisation group scale $k$  in terms of an operator trace over the full propagator times the scale derivative of the cutoff itself $(t=\ln {k}/{\Lambda})$. 
The flow derives from a path-integral representation of the theory with partition function $
Z_k[J]=\int D\varphi\exp(-S[\varphi]-\Delta S_k[\varphi]-\varphi\cdot J)$
where $S$ denotes the classical action, $J$ an external current, and the Wilsonian cutoff term  
at momentum scale $k$ is given by
\beq\label{deltaS}
\Delta S_k[\varphi]=\frac12\int\,\frac{d^dq}{(2\pi)^d}\,\varphi(-q)\, R_k(q^2)\,\varphi(q)\,.
\eeq
The function $R_k$ obeys the limits $R_k(q^2)>0$ for $q^2/k^2\to 0$ and $R_k(q^2)\to 0$ for $k^2/q^2\to 0$ to make sure it acts as an IR momentum cutoff \cite{Litim:2001up,Litim:2000ci,Litim:2001fd}. Optimised choices for the regulator term \cite{Litim:2000ci,Litim:2001up,Litim:2001fd,Litim:2010tt} allow for analytic flows and an improved convergence of systematic approximations \cite{Litim:2005us}. 
 The partition function
falls back onto the full physical theory in the limit $k\to 0$.
 As such, the flow \eq{FRG} interpolates between  a microscopic  (classical) theory $(k\to\infty)$
and the full quantum effective action $\Gamma$ $(k\to 0)$,
\beq\label{k0}
\lim _{k\to 0}\Gamma_k=\Gamma\,.
\eeq
At weak coupling, iterative solutions generate the perturbative loop expansion \cite{Litim:2001ky,Litim:2002xm}. Also, the flow \eq{FRG} relates to the Polchinski flow \eq{floweqn} by means of a Legendre transformation. 

In the sequel, we are interested in $O(N)$ symmetric scalar field theory in $d$ euclidean dimensions to leading order in the derivative expansion \cite{Berges:2000ew}, with the effective potential  $W_{k} (\phi^a \phi_a)$.
The anomalous dimension of the field vanishes, the wavefunction does not get renormalised  
and the approximation becomes exact  \cite{Ma:1973zu}, and exactly soluble  
\cite{Ma:1973zu,Tetradis:1995br,Litim:1995ex,DAttanasio:1997yph}.  Within the local potential approximation, the 
massless  cutoff propagator is given by
\beq\label{PkERG}
P_k(q)=\frac{1}{q^2+R_k(q^2)}\,.
\eeq
Here, and unlike in \eq{PkPol}, the Wilsonian scale $k$ takes the role of an IR cutoff.
Introducing dimensionless variables
$w(z)=W/k^d$ together with
$z=\frac12 \phi^2\,k^{2-d}$
 one finds the flow equation for the dimensionless potential
\begin{equation}\label{du}
\partial_t w = -dw+(d-2)z w^\prime +(N-1)I[w']+I[w'+2z w'']
\end{equation}
in $d$ euclidean dimensions. The functions $I[x]$, also known as ``threshold functions'' \cite{Wetterich:1992yh}, relate to the loop integral in \eq{FRG} and are given by 
\begin{equation}\label{Igen}
I[x]=\frac12\, k^{-d}\int \frac{d^dq}{(2\pi)^d}\, \frac{\partial_t R_k(q^2)}{q^2+R_k+x\, k^2}\,.
\end{equation}
In given approximations, the RG flows, their fixed points, and universal observables genuinely depend on the regulator \eq{deltaS}, \eq{Igen}. It is well known that suitably adapted, optimised choices can improve convergence and stability of approximations \cite{Litim:2001fd,Litim:2001dt,Litim:2002qn,Litim:2002cf,Litim:2010tt}.
An important regulator function is the ``optimised'' one given by \cite{Litim:2002cf,Litim:2001up,Litim:2000ci} 
\begin{equation}\label{opt}
R_{\rm opt}(q^2)=q^2 \left(1+r_{\rm opt}(q^2/k^2)\right)\,,\quad r_{\rm opt}(y)=\left(\01y-1\right)\theta(1-y)\,,
\end{equation}
which obeys a simple criterion for stability \cite{Litim:2000ci}. In the local potential approximation, the function \eq{opt} can be understood as the ``convex hull''  of generic regulators \cite{Litim:2000ci,Litim:2001up}. As such, it leads to very good convergence and stable results for physical observables \cite{Litim:2007jb}.
The choice \eq{opt} also ensures exact equivalence with findings from the Polchinski flow \cite{Litim:2002cf,Litim:2003kf,Litim:2005us,Morris:2005ck,Litim:2007jb,Bervillier:2007rc}.
The integration in \eq{Igen} with  \eq{opt} gives
\begin{equation}\label{I}
I_{\rm opt}[x]=\frac{2}{d\,L_d}\frac{1}{1+x}\,.
\end{equation}
The numerical factor $L_d=(4\pi)^{d/2}\Gamma(d/2)$ denotes the $d$-dimensional loop factor \cite{Litim:2016hlb}. After a straightforward rescaling 
we obtain Litim's version for the optimised functional flow   { \cite{Litim:2001up,Litim:2002cf}
\begin{equation}\label{duprime}
\partial_tw=-d w+(d-2)zw'+(N-1)\frac{1}{1+w'}+\frac{1}{1+w'+2zw''}\,,
\end{equation}
modulo an irrelevant constant term.  We have  rescaled the field and the potential as $u\to u/A$ and $\rho\to \rho/A$ with $A=2/(d\,L_d)$.
Apart from the factor $2/d$ (which originates due to the cutoff profile function), all couplings are now measured in units of 
perturbative loop factors consistent with  naive dimensional analysis.

\subsection{Duality}\label{duality}
\label{appendix}
Polchinski's and Wetterich's equation are both exact renormalisation group flows, though different on a  conceptual and practical level \cite{Berges:2000ew,Litim:2005us,Litim:2011cp}. Specifically, the former is based on a UV, while the latter uses an IR cutoff. Also, both flows have inequivalent derivative expansions and scheme dependencies, and the relevant non-linerarities appear in substantially different manners \cite{Litim:2005us}. 
Interestingly though, it has been established that the RG flow equations can be mapped  onto each other, both on the level of the full flows and on the level of specific approximations such as the leading order of the derivative expansion adopted here. In particular, both flows give {\it identical} scaling exponents within a local potential approximation (LPA) {\it if and only if}  
 the optimised cutoff \eq{opt} is adopted for the Wetterich flow \cite{Litim:2001up,Litim:2002cf,Litim:2002cf,Litim:2002qn,Litim:2005us,Bervillier:2007rc,Litim:2007jb}. By now, their  equivalence \cite{Litim:2002qn,Litim:2005us} is additionally supported by an explicit map in LPA  \cite{Morris:2005ck,Bervillier:2007rc,Bridle:2016nsu}.

For later convenience we summarise the main steps.
For an $O(N)$ symmetric scalar field theory  to  leading order in the derivative expansion, the Polchinski flow \eq{incneqn} and Litim's version  of the Wetterich flow   \eq{duprime} 
 take the form
\begin{equation}\label{both}
\left\{\begin{array}{rl}
\partial_tu&=-d u +(d-2)\rho u'+2\rho(u')^2-(N-1) u'-(u'+2\rho u'')\\[1ex]
\partial_tw&
\displaystyle
=-d w+(d-2)zw'+(N-1)\left(\frac{1}{1+w'}-1\right)+\left(\frac{1}{1+w'+2zw''}-1\right)\,,
\end{array}
\right.
\end{equation}
respectively, where we have used $u=u(\rho,t)$ and $w=w(z,t)$, with  $\rho=\half\varphi_a\varphi_a$ and $z=\half\phi_a\phi_a$. Both flows are normalised to vanish for vanishing potentials $u$ or $w$. In the local potential  approximation, the fields are space-time independent and both of \eq{both} are related by a Legendre ``duality'' transformation \cite{Morris:2005ck}
\begin{equation}\label{legpot}
u(\rho,t)=w(z,t)+\half(\phi_a-\varphi_a)^2 .
\end{equation}
It follows that the flows on the LHS of \eq{both} are the same, 
$\partial_t u=\partial_t w$
as the fields $\varphi_a$ and $\phi_a$ are $t$-independent. Moreover, the relations
\begin{equation}\label{legphi}
\phi_a-\varphi_a=-\varphi_a u'=-\phi_aw'
\end{equation}
imply that $\varphi_a$ and $\phi_a$ point into the same direction, therefore
\begin{equation}\label{leguprel}
\sqrt{\frac{z}{\rho}}=1-u'=\frac{1}{1+w'}\,.
\end{equation}
The result is pivotal for linking  the two equations, giving the key relation 
\beq\label{key}
w'=\frac{u'}{1-u'}\,,
\eeq
Moreover, squaring \eq{leguprel} 
gives us the substitutions for the field variables
\begin{equation}\label{legfield}
z=\rho (1-u')^2\,.
\end{equation}
We also learn  from \eq{legphi} that $u'\sqrt{\rho}=w'\sqrt{z}$, which, with \eq{leguprel},   leads to $z w'-\rho u' =-\rho (u')^2$, while the square of \eq{legphi}  implies $u-w=\rho(u')^2$. Taken together, this provides us with the important relation
\begin{equation}\label{can}
-d w +(d-2)z w'=-d u+(d-2)\rho u'-2\rho(u')^2\,.
\end{equation}
Notice that \eq{can} appears to map the terms with canonical scaling of the Wetterich flow into the corresponding canonical terms for the Polchinski flow, plus the non-linear term $-2\rho(u')^2$. Finally by differentiating (\ref{leguprel}) with respect to $\rho$ and differentiating the inverse by $z$ we find the relation
\begin{equation}\label{inv}
1-u'-2\rho u''=\frac{1}{1+w'+2zw''}.
\end{equation}
In combination, \eq{leguprel}, \eq{legfield},  \eq{can} and \eq{inv}  show the exact term-by-term equivalence of the two flow equations in \eq{both} including the irrelevant and  field-independent vacuum term $-N$ \cite{Morris:2005ck,Bridle:2016nsu} . 

\begin{figure*}[t]
	\centering
	\includegraphics[scale=0.45]{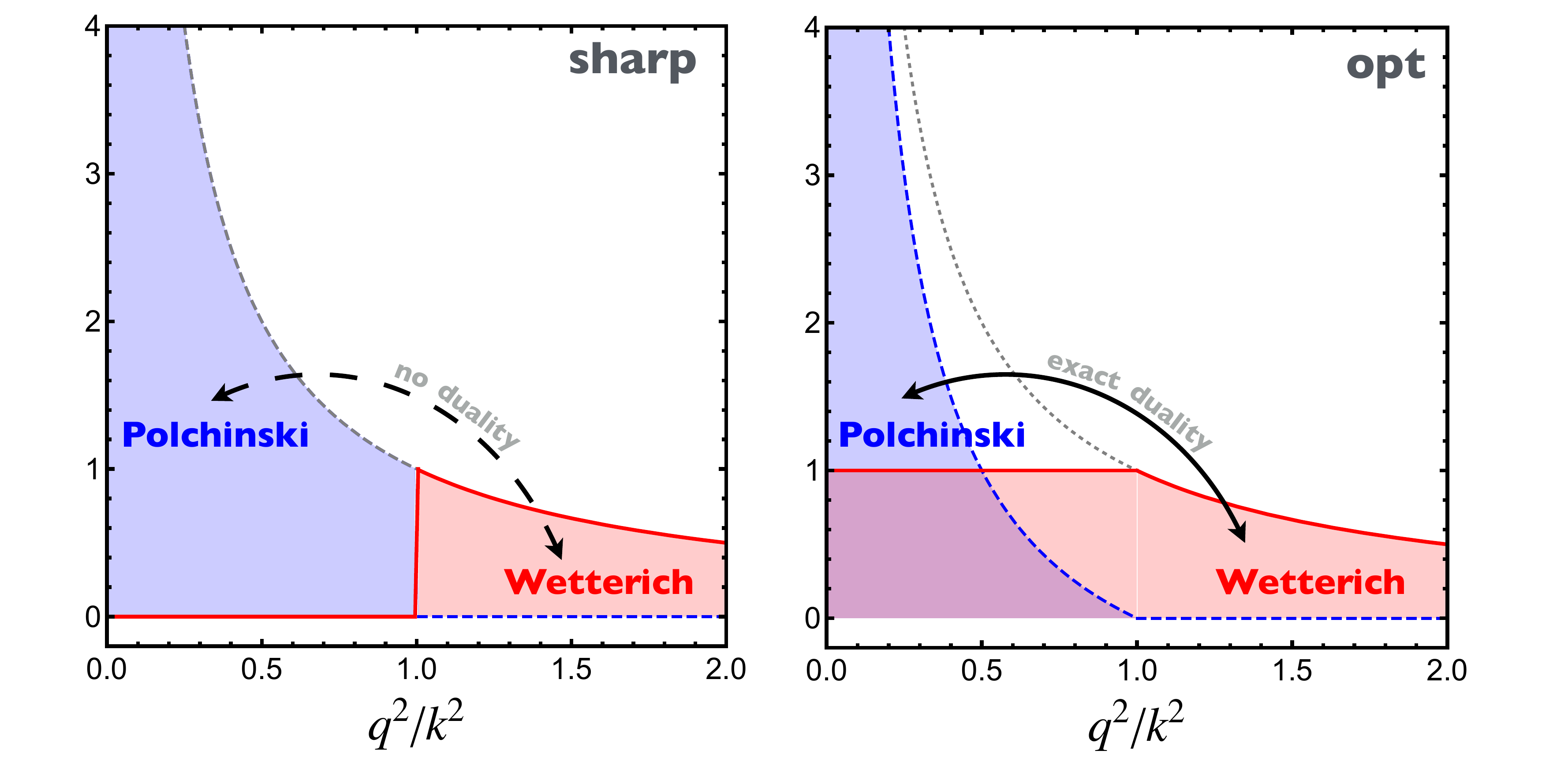}
	\caption{Polchinski-Wetterich duality and propagator sum rule. Shown are different UV-IR decompositions of the massless propagator into a Polchinski part $P_{{}_{\rm UV}}$ where $k$ takes the role of a UV cutoff (blue), and a Wetterich part $P_{{}_{\rm IR}}$ where $k$ acts as an IR cutoff (red).  Only for the decomposition \eq{sum},  the respective Polchinski and Wetterich flows \eq{both} are exact duals of each other (right panel). Dualities cannot be found for any other  decomposition \eq{generic} including the sharp momentum cutoff  (left panel).}
	\label{pPropDuality}
\end{figure*}

Some aspects of the duality \eq{both} can  be appreciated with the help of a sum rule  \cite{Bridle:2016nsu}. For an optimised pair of UV and IR regulators \eq{Kopt}, \eq{opt}, the free massless  propagator 
$P_{{}_{\rm LPA}}=1/q^2$ can be decomposed uniquely into a UV  part $P_{{}_{\rm UV}}(q,k)=K_{\rm opt}/{q^2}$  where $k$ takes the role of a UV cutoff \eq{Kopt}, and an IR  part $P_{{}_{\rm IR}}(q,k)=1/(q^2+R_{\rm opt})$  for which $k$ takes the role of an IR cutoff, \eq{opt}. The former arises in the Polchinski flow, while the latter is employed in the optimised Wetterich flow. Irrespective of the value for $k$, we  observe the exact sum rule $P_{{}_{\rm LPA}}=P_{{}_{\rm UV}}+P_{{}_{\rm IR}}$ between either of these, $i.e.$
\beq\label{sum}
P_{{}_{\rm UV}}+P_{{}_{\rm IR}}\equiv \frac{K_{\rm opt}}{q^2}+\frac{1}{q^2+R_{\rm opt}}=\frac{1}{q^2}
\eeq
At infinite $N$, the anomalous dimension of the scalar field vanishes identically. Thereby the local potential approximation of the Wetterich  flow becomes exact and  the duality \eq{sum} holds true non-perturbatively.
In principle,  a unique Polchinski regulator $K$ could be defined for any Wetterich regulator $R$ by {\it imposing} the sum rule \eq{sum}, leading to
\beq\label{generic}
K(q^2/k^2)=\frac{R_k(q^2)}{q^2+R_k(q^2)}\,.
\eeq
In this spirit, the seminal sharp cutoff limit
$R_{\rm sharp}=\lim_{a\to \infty} a\cdot \theta(k^2-q^2)$ for the Wetterich flow would correspond to a simple step function cutoff $K(y)=\theta(1-y)$ for  the Polchinski flow.  However, it is only for the "optimised" regulators    $R=R_{\rm opt}$   \eq{opt} and $K=K_{\rm opt}$  \eq{Kopt} that the sum rule decomposition \eq{sum}  leads to  Polchinski and Wetterich flows \eq{both} which are {\it exact duals}  of each other under the Legendre  transformation \eq{legpot}, for any RG scale. These results  are further  illustrated in Fig.~\ref{pPropDuality}, which displays the massless propagator in units of $k$ and its decompositions into Polchinski and Wetterich propagators for, exemplarily, the sharp  and the optimised cutoff. Only the latter entails exact duality.

\section{\bf Running couplings}\label{RC}
In this section, we provide the local and global beta functions for all couplings within the Polchinski interaction potential and their solutions for all RG scales.
\subsection{Local flows}\label{local}
We expand the Polchinski flow  around the potential minimum by making a polynomial approximation of the potential
\begin{equation}\label{A}
u=\sum\limits_{n=2}^{\infty}\frac{\lambda_n(t)}{n!}(\rho-\kappa(t))^n \,,
\end{equation}
where we have defined $\kappa$ as the potential minimum with $u'(\kappa)=0$ \cite{Aoki:1996fn}. The flow for all couplings $\lambda_n$ obtains from \eq{polsca} by suitable derivatives. We find
\beq\label{lambdan}
\partial_t\lambda_n=(n-3)\lambda_n +f_n(\kappa,\lambda_{i\le n})\,.
\eeq
The first term arises from the canonical mass dimension of the coupling $\lambda_n$. The second term, $f_n$, arises due to  quantum fluctuations. Here, the functions $f_n$ are polynomials of their arguments. 
The flow for the vacuum expectation value $\kappa$ is found by exploiting $du'(\kappa,t)/dt=0$ with \eq{polsca}.  Using the techniques of \cite{Litim:2016hlb,Juttner:2017cpr}, all polynomial fixed points can be found recursively from \eq{lambdan} by solving $\partial_t\lambda_n=0$ for the highest coupling, which is $\lambda_n$. 

As an aside, we note that exact RG flows $\partial_t\lambda_n$ depend, ordinarily, on couplings up to including $\lambda_{n+2}$ due to the second order nature of the underlying functional RG flow \eq{floweqn}, \eq{FRG}.
Here, this dependence is reduced to $\lambda_{n+1}$ due to the large-$N$ limit, \eq{both}. An additional reduction to $\lambda_{n}$ arises in \eq{lambdan} due to the specific expansion point in \eq{A}, the running potential minimum.

Using the notation $\lambda\equiv\lambda_2$ and $\tau\equiv \lambda_3$, we find the first few flow equations from \eq{lambdan} as
\beq
\label{lock}
\begin{array}{rcl}
\partial_t\kappa&=&1-\kappa\\ [1ex]
\partial_t\lambda&=&-\lambda(1-4\kappa\lambda)
\\[1ex]
\partial_t\tau&=&12\lambda(\lambda+\kappa\tau) \\[1ex]
\partial_t\lambda_4&=&\lambda_4 +12\kappa\,\tau^2+ 16 \lambda(3\tau+\kappa\,\lambda_4) \\
&\vdots&
\end{array}
\eeq
and similarly for the higher order couplings.  Notice that the flow of the vacuum expectation value ``factorises'' as it is entirely independent  from all other couplings. The flow of the quartic coupling $\lambda$ only depends on the quartic itself, and on $\kappa$. This pattern extends to higher order: the flow for $\lambda_n$ depends on $\kappa$ and couplings $\lambda_m$ with $m\le n$.  
The fixed point solution for the dimensionless vacuum expectation value (VEV) $\kappa$ when $\partial_t\kappa=0$ is
\begin{equation}
\kappa_*=1\,.
\end{equation}
Let us introduce $\kappa_\Lambda$, $\lambda_\Lambda$ and $\tau_\Lambda$ to denote the initial values of couplings at $k=\Lambda$. 
The analytical solution for the flow of the potential minimum is 
\begin{equation}
\kappa(t)=\kappa_*+c_\kappa e^{-t}
\end{equation}
with integration constant $c_\kappa=\kappa_\Lambda-1$. We observe  that $\kappa=\kappa_*$ is a UV attractive fixed point. Consequently,  the VEV  is a relevant operator. Now considering the quartic coupling the fixed point solutions are
\begin{equation}\label{lambdaWF}
\lambda_*=0 \;\;\;\;\; \text{or} \;\;\;\;\; \lambda_*=\frac{1}{4}.
\end{equation}
The analytical solution for the running of the quartic coupling is
\begin{equation}\label{lambda}
\lambda(t)=\frac{e^{t}}{4e^{t}+2c_\kappa+c_\lambda e^{2t}}
\end{equation}
with integration constant $c_\lambda=1/\lambda_\Lambda-2c_\kappa-4$. We see that the fixed point $\lambda_*=0$ corresponds to a UV attractive fixed point. Conversely, $\lambda_*=1/4$ corresponds to an IR attractive fixed point,  the seminal Wilson-Fisher fixed point. Finally for the sextic coupling we find the fixed point values corresponding to $\lambda_*=0$ and $\lambda_*=1/4$ respectively as
\begin{equation}
\tau_*=\tau \;\;\;\;\; \text{or} \;\;\;\;\; \tau_*=-\frac{1}{4}.
\end{equation}
We note that the flow for $\tau$ vanishes identically for vanishing quartic coupling, implying that $\tau$ becomes an exactly marginal coupling which characterises a line of interacting fixed points.
The analytical solution for the running of $\tau$ is
\begin{equation}
\tau(t)=\frac{e^{2t}(-6c_\kappa+e^{t}(-16-6e^tc_\lambda+e^{3t}c_\tau))}{(2c_\kappa+e^t(4+e^tc_\lambda))^3}
\end{equation}
with integration constant 
\bea
c_\tau&=&16+64\tau_\Lambda+8c_\lambda^2\tau_\Lambda+c_\lambda^3\tau_\Lambda
+12c_\kappa^2(4+c_\lambda)\tau_\Lambda
+6c_\lambda(1+8\tau_\Lambda)
+6c_\kappa(1+(4+c_\lambda)^2\tau_\Lambda)\,.
\eea
All further couplings $\lambda_n$ for $n>3$ can be obtained recursively in terms of $\kappa_*$, $\lambda_*$ and $\tau_*$.
 
 We conclude that two types of interacting fixed points arise. Firstly, when the quartic coupling vanishes we have the line of  fixed points
\begin{equation}\label{loctcfp}
(\kappa_*\lambda_*,\tau_*)=(1, 0,\tau)\,,
\end{equation}
where $\tau$ serves as a free parameter which remains unrenormalised non-perturbatively. This line of UV fixed points displays asymptotic freedom at the Gaussian fixed point. Once $\tau\neq 0$, the theory displays asymptotic safety. Secondly, we also find a unique isolated fixed point -- the Wilson-Fisher fixed point -- with  
\begin{equation}\label{locwffp}
(\kappa_*,\lambda_*,\tau_*)=\left(1,\frac{1}{4},-\frac{1}{4}\right)\,.
\end{equation}
The negative sign of the sextic coupling for the IR fixed point is an interesting consequence of the Polchinski equation and suggests that the potential might be unbounded from below. However, we will see that this is not the case, owing to the higher order couplings 
$\lambda_n$.

\subsection{Global flows}
Returning to the Polchinski flow  (\ref{polsca}) we now obtain its general, global solutions via the method of characteristics \cite{Litim:1995ex,Tetradis:1995br}.
The procedure leads to full, non-perturbative solution for all fields without 
making any assumptions about the form of the solution. We will find, however, that physical fixed point solutions genuinely falls back onto the local flows under suitable polynomial approximations.  Starting from \eq{polsca}, we are lead to a set of ordinary differential equations for the characteristics,
\begin{eqnarray}\label{char1}
\frac{du'}{dt}&=&-2u'(1-u')\\
\label{char2}
\frac{d\rho}{du'}&=&\frac{\rho}{2u'}\frac{1+4u'}{1-u'}-\frac{1}{2u'(1-u')}\,.
\end{eqnarray}
The first characteristics is integrated to give $e^{2t}\,{u'}/(1-u')=$~const. For the second one we must consider the cases
$u'\geq0$ and $u'<0$ separately.
In combination, we find the general solution as
\begin{equation}\label{gpos}
\frac{\rho-1}{\sqrt{|u'|}}{(1-u')^{5/2}}-F(u')=G\left(e^{2t}\frac{u'}{1-u'}\right)\,.
\end{equation}
The function $G$ is unspecified, and solely determined through initial conditions for the derivative of the interaction potential $u'$ at some reference energy scale $k=\Lambda$. The function $F$ arises from the second characteristics and is given by
\begin{equation}\label{F}
F(u')=
\left\{
\begin{array}{lcl}
\displaystyle
\ \ {\sqrt{1-u'}}\ {\sqrt{u'}}\ \left(\frac52-u'\right)+\frac{3}{2}\arcsin\sqrt{u'}
&\text{\ \ for\ \ }&1>u'>0\\[3ex]
\displaystyle
{-\sqrt{1-u'}}{\sqrt{-u'}}\left(\frac52-u'\right)-\frac{3}{2}\arcsinh\sqrt{-u'}
&\text{\ \ for\ \ }&u'<0
\end{array}
\right.
\end{equation}
which is monotonous in $x$ with $F(x)\ge 0$ for $x\in[0,1]$, $F(x)< 0$ for $x\in[-\infty,0)$,  $F(0)=0$ and $F(1)=\frac{3\pi}{4}$. Analytic continuity of $F$ follows from  $\arcsin z=-i\ln(iz+\sqrt{1-z^2})$ and  $\arcsinh z =\ln(z+\sqrt{1+z^2})$ in the complex $z$ plane, together with $i\sqrt{z}=\sqrt{-z}$ when $z$ changes sign.  We also observe that   $u'$ takes values within the range $(-\infty,1)$. An analytic continuation   beyond $u'=1$ is not available in general except for specific choices for $G$.
This structure is consistent with the  mapping onto the dual  flow \eq{both}, also using \eq{key} 
for $w'$ in the range $w'\in(-1,\infty)$.
We also note that using the Legendre transform relation for the first derivative of the potential (\ref{leguprel}), the field transform relation (\ref{legfield}) and the trigonometric identity $\arcsin(x)=\arctan({x}/{\sqrt{1-x^2}})$ the global solution (\ref{gpos}), \eq{F} maps exactly onto the global solutions given in  \cite{Litim:1995ex,Marchais:2017jqc} for the optimised Wetterich flow at infinite $N$ \cite{Litim:2001up,Litim:2002cf}.

\subsection{Interacting fixed points}
Fixed point solutions $u'_*(\rho)$ are scale independent. In view of \eq{gpos} scale independence implies that the function $G$  reduces to a $t$-independent constant. We define the constant such that we always have a real parameter $c$. 
 We then find that all continuous and differentiable fixed point solutions must obey the relation
\begin{equation}\label{solution}
\left|\frac{\rho-1}{\sqrt{|u'|}}\, (1-u')^{5/2} -F(u')\right|=
\left\{
\begin{array}{cl}
c&
\ (u'\ge 0)\\[3ex]
-c&\ (u'< 0) 
\end{array}
\right.
\end{equation}
where the parameter $c$ can take any positive (for $u'>0)$ or negative  (for $u'<0)$  real value, and $F$ is given in \eq{F}.
\begin{figure}[t]
	\centering
	\includegraphics[scale=0.25]{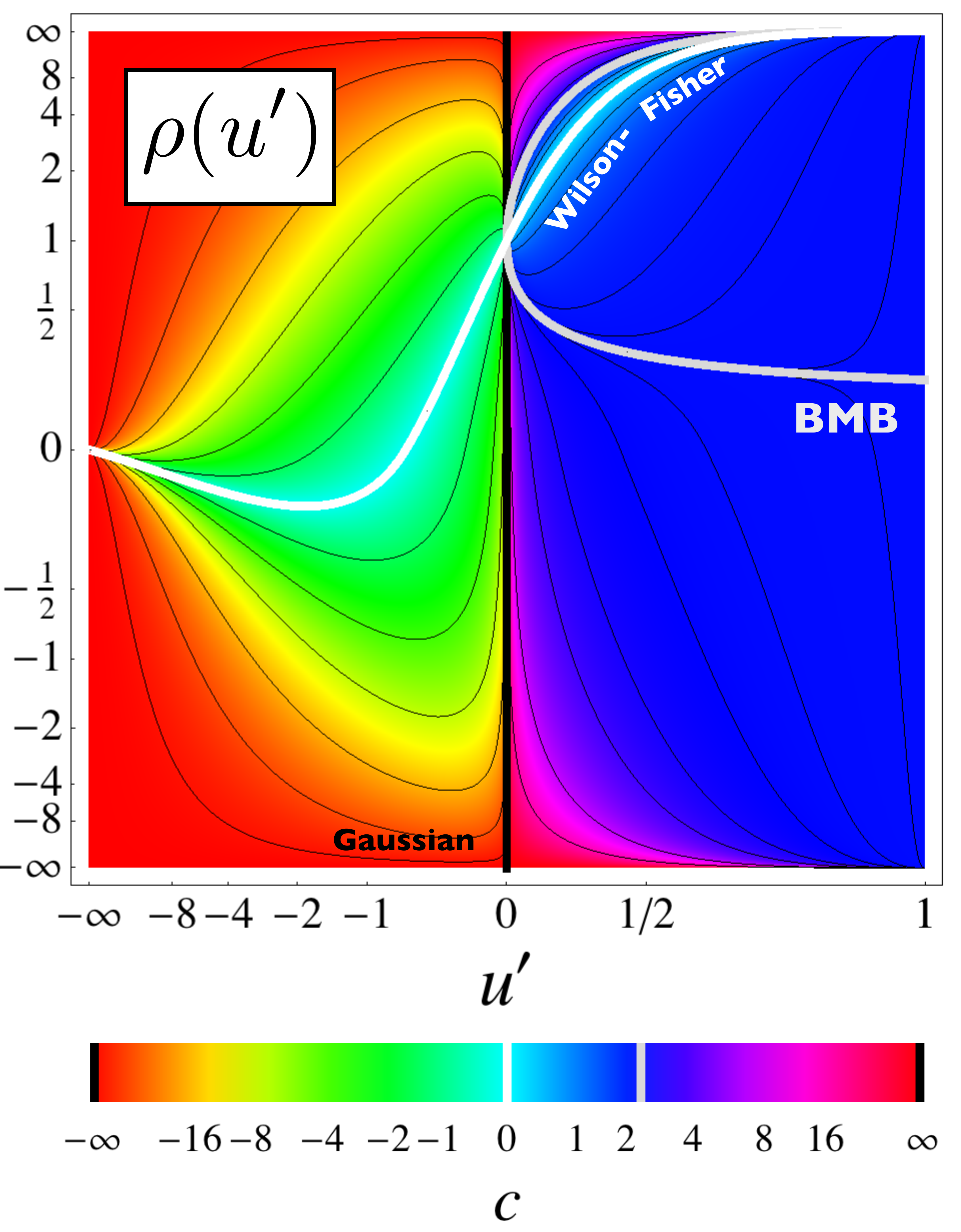}
	\caption{Global fixed points $\rho(u')$ of the Polchinski equation in the large-$N$ limit for all fields $-\infty\leq\rho\leq\infty$ and potentials $-\infty\leq u'\leq1$, in dependence on the free parameter $c$ and color-coded according to the legend. Axes are rescaled as $\rho\rightarrow{\rho}/(1+|\rho|)$ and $u'\rightarrow {u'}/(2-u')$ for better display. Thick lines indicate the Gaussian fixed point (black, $|c|=\infty$), the Wilson-Fisher fixed point (white, $c=0$), and the Bardeen-Moshe-Bander (BMB) fixed point (grey, $c=3\pi/4$), thin black lines for constant $c$ are added to guide the eye.}
	\label{largeNdense}
\end{figure}
Note that the absolute value in \eq{solution} is essential to ensure analyticity around the saddle point 
\begin{equation}\label{saddle}
(\rho,u')=(1,0) \,.
\end{equation}
For any $c$, any integral  curve will pass through \eq{saddle}, which arises from the solution of (\ref{polsca}) at the fixed point.
Integral curves with $u'>1$  exist as well. None of  these (bar one, see Sec.~\ref{BMB}) can be written as \eq{solution}
because $u'=0$ cannot be reached. Under the duality map to the effective potential \eq{both}, \eq{key} and \eq{legfield},  stable interaction potentials with $u'>1$ and $\rho>0$ are mapped onto unstable effective potentials with $w'<0$ and and $z>0$, which is why these solutions are considered unphysical and are not investigated any further.

In Fig.~\ref{largeNdense} we show the global fixed point solutions in the form $\rho=\rho(u')$ as given in \eq{solution} and as a function of the free parameter $c\in (-\infty,\infty)$. Axes have been rescaled as $\rho\rightarrow{\rho}/({1+|\rho|})$ and $u'\rightarrow {u'}/({2-u'})$  for better display. All curves with constant $c$ are color-coded according to the legend. Thin black curves with constant $c$ have been added to guide the eye. 

Of particular interest are those fixed point solutions which are unique (single-valued), and which exist for all physical fields $\rho\ge 0$. 
A few distinct cases are worth being highlighted, see Fig.~\ref{largeNdense}. Firstly, the limit
\beq\label{cGauss}
|c|\to \infty
\eeq
entails that $u'$ vanishes identically for all fields. This corresponds to the Gaussian fixed point (full black line), which  trivially exists for all fields. The non-interacting fixed point is an UV fixed point of the theory, corresponding to ``asymptotic freedom''. 
Secondly, the case
\beq\label{cWF}
c=0
\eeq
also leads to a unique fixed point solution $u'(\rho)$ for all $\rho\ge 0$. We note that $u'$ changes sign at $\rho=1$ implying that it relates to the minimum of the fixed point potential.  As we will establish shortly, this case corresponds to an IR fixed point, the unique Wilson-Fisher fixed point (full white line) of the theory. Thirdly, within the range
\beq\label{cTri}
c_{\rm crit}<c<\infty
\eeq
with $c_{\rm crit}$ given in \eq{cBMB} below, we observe from Fig.~\ref{largeNdense}  a family of integral curves which exist for all $\rho\ge 0$. However, and much unlike the Wilson-Fisher fixed point, we find that $u'\ge 0$ for all of these, for all fields. For this reason, $\rho=1$ corresponds to a saddle point rather than a minimum of the fixed point potential. As we will show in more detail below, these fixed points are UV fixed points where the theory remains interacting at shortest distances. This phenomenon is also known as ``asymptotic safety''.
Finally, the limiting case where $c=c_{\rm crit}$ with
\beq\label{cBMB}
c_{\rm crit}\equiv\frac{3\pi}{4}
\eeq
is of particular interest. For any $c<c_{\rm crit}$, the integral curves do not cover the entire range of physical fields $\rho\ge 0$. For this reason, they are discarded as physical fixed point candidates. Intriguingly though, the borderline case \eq{cBMB} (full gray line) still offers a physical limit with novel features related to the spontaneous breaking of scale invariance first discovered by Bardeen, Moshe and Bander. Below, we refer to models with $c>c_{\rm crit}$ $(c=c_{\rm crit})$ [$c<c_{\rm crit}$] as weakly (critically) [strongly] coupled. 

In the remaining parts of the paper we concentrate on each of these settings \eq{solution}, for all $c$, including the phase diagram of the full theory. 

\section{\bf Infrared fixed point}\label{IR}
In this Section, we discuss the Wilson-Fisher fixed point solution of the Polchinski equation for all fields including in the complex plane. We also investigate radii of convergence for expansions, singularities in field space, and universal critical exponents.
\subsection{Wilson-Fisher fixed point}
The isolated Wilson-Fisher fixed point arises for the parameter choice \eq{cWF}. In this case, the relation \eq{solution} takes the form
\begin{equation}\label{rhowf}
\rho_{\rm WF}(u')=1+H(u')
\eeq
where we have introduced the auxiliary function
\beq\label{H}
H(x)=
\frac{\sqrt{|x|}\,F(x)}{(1-x)^{5/2}}
\end{equation}
with $F(x)$ given in \eq{F}. The function $H(x)$ contains all the relevant information. It is analytical in $x$ and so is  the expression \eq{rhowf}  in $u'$ across $u'=0$. It obeys $H(x)>0$ for $1>x>0$ and $H(x)<0$  for $x<0$. From Fig.~\ref{largeNdense} we observe that $u'$ takes values 
up to $u'\to 1$, corresponding to the regime where fields are large. For this reason we are interested in $H(x)$ in the small and large field regions, where it has the following expansions 
\begin{eqnarray}\label{Hsmall}
H(x)&=&4x+8x^2+\frac{64}{5}x^3+\mathcal{O}(x^4)\,,\\
\label{Hlarge}
H(1-x)&=&\frac{3\pi}{4}\left(\frac{1}{x^{5/2}}
-\frac{1}{2x^{3/2}}
-\frac{1}{8x^{1/2}}
-
\frac{16}{15\pi}
\right)
+{\cal O}(\sqrt{x})
\end{eqnarray}
for small $x\ll 1$. 
The Taylor expansion of \eq{rhowf}  for small argument can be read off from \eq{Hsmall}.
Inverting the Taylor expansion order by order, the corresponding expansion around (\ref{saddle}) in field space takes the form
\begin{equation}\label{wfana}
u'=\frac{1}{4}(\rho-1)-\frac{1}{8}(\rho-1)^2+\frac{3}{40}(\rho-1)^3+\cdots
\end{equation}
up to terms of order $\mathcal{O}[(\rho-1)^4]$  in the fields. The quartic and sextic couplings at the minimum agree with the local solution for the IR attractive fixed point (\ref{locwffp}), which identifies it as the Wilson-Fisher fixed point. 
Taylor-expanding the solution \eq{rhowf} around $u'\le 1$ with the help of \eq{Hlarge}  we find that
\bea
\rho&=&\frac{3\pi}{4(1-u')^{5/2}}
-\frac{3\pi}{8(1-u')^{3/2}}
-\frac{3\pi}{32(1-u')^{1/2}}
+\frac{1}{5}
+{\cal O}(\sqrt{1-u'})\,,
\eea
showing that $\rho$ grows large algebraically in $1/(1-u')$ in the limit $u'\to 1$,   with the index $\frac 52$. Inverting the large-field expansion we find that the derivative of the fixed point potential asymptotes as
\beq\label{asymWF}
u'(\rho)=1-\left(\frac{3\pi}{4\rho}\right)^{2/5}+\text{subleading}
\eeq
towards unity with increasing $\rho$. 

In Fig.~\ref{pWFcomplex} we show the Wilson-Fisher fixed point in the entire complexified field space. The left (right) panel displays the real (imaginary) part of $\rho$ as a function of $u'$ in the entire complex $u'$ plane. Along the real $\rho$ axis the fixed point (full red line) cannot be extended beyond the singularity at $u'=1$ without acquiring an imaginary part (dashed red line). For asymptotically large $|u'|$, we always find that $\rho\to 0$.

\begin{figure*}[t]
	\centering
	\includegraphics[scale=0.4]{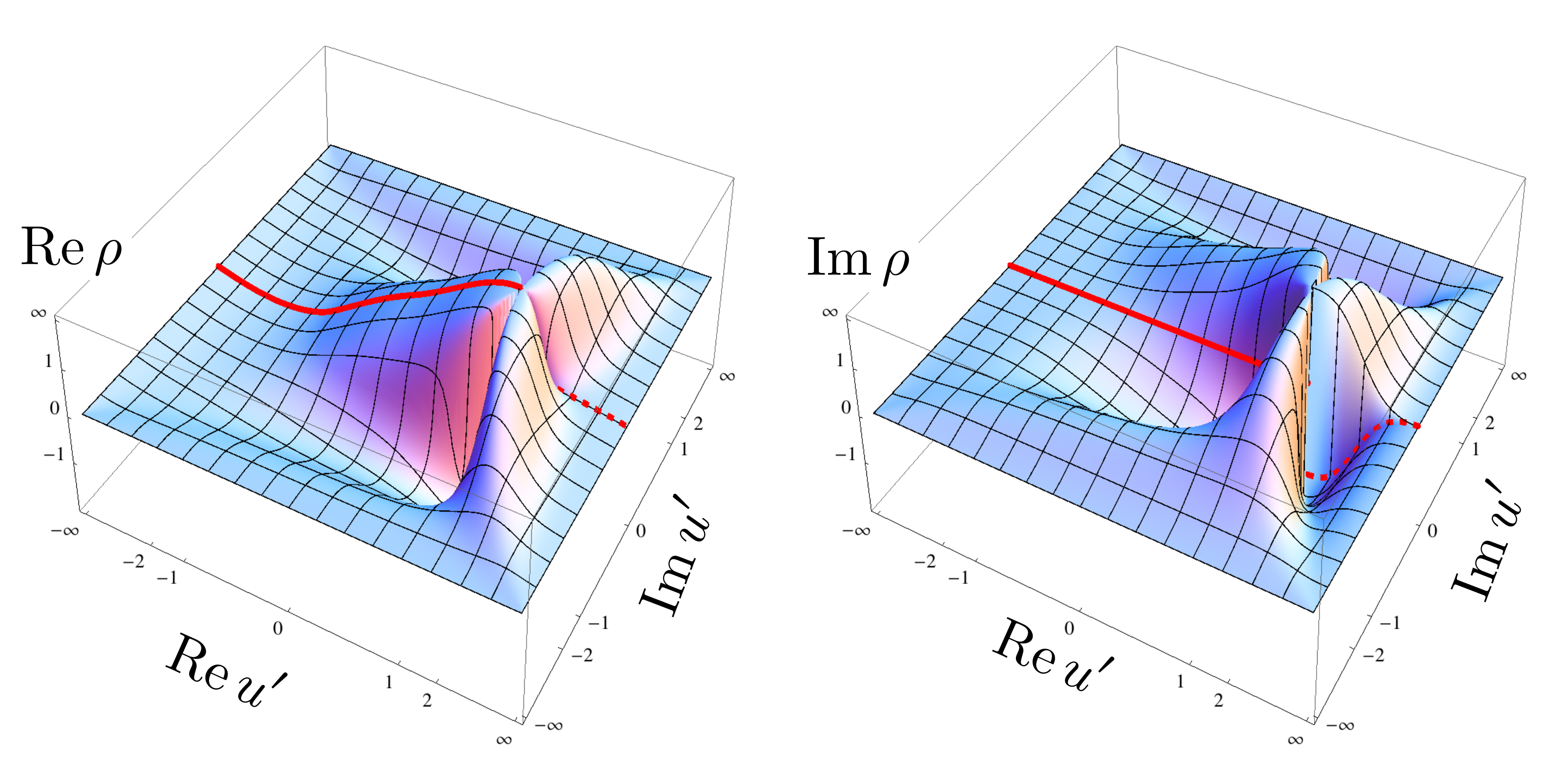}
	\caption{Global Wilson-Fisher fixed point from the Polchinski flow in the entire complex field plane, showing the real (left panel) and imaginary (right panel) part of the field variable $\rho=\frac12 \phi^a\phi^a/k$ as a function of $u'$ in the entire complex  $u'$ plane. Axes are rescaled as $X\to X/(1+|X|)$ for better display. We observe a pole at $u'=1$. Along the real $\rho$ axis, we observe that the Wilson-Fisher fixed point (full red line) cannot be extended beyond the pole at $u'=1$ without acquiring an imaginary part (dashed red line).}
	\label{pWFcomplex}
\end{figure*}

\subsection{Bootstrap for scaling exponents}

Next, we compute the scaling exponents at the Wilson-Fisher fixed point. 
Within a polynomial approximation \eq{A}, the fixed point search strategy has the form of a bootstrap \cite{Falls:2013bv}. A finite order polynomial expansion is based on the tacit assumption that interaction terms with higher canonical mass dimension correspond to less relevant operators at an interacting fixed point, meaning that universal eigenvalues must increase with increasing canonical mass dimension of the highest retained invariant $ \propto \lambda_n\int d^dx \rho^n$ in the action. This assertion, however, must be checked a posteriori, which turns this procedure into a bootstrap \cite{Falls:2013bv}.\footnote{For applications of the bootstrap in $4d$ quantum gravity, see \cite{Falls:2014tra,Falls:2017lst}.} Following  \cite{Falls:2013bv}, we calculate universal scaling exponents describing the behaviour of physical observables close to a fixed point. We calculate this via the stability matrix
\begin{equation}
\partial_t \vec{v}_i=\mathcal{M}\vec{v}_i=\theta_i\vec{v}_i\,.
\end{equation}
where $\mathcal{M}_{ij}=d\beta_i/d \lambda_j|_{\vec{\lambda}=\vec{\lambda}_*}$, and $\lambda_j$ are the coupling constants and $\lambda_{j*}$ their fixed point values. The   universal scaling exponents $\theta_i$ of (small) eigenperturbation $\vec{v}_i$ with components  ${v}_i=\lambda_{i}-\lambda_{i*}$ are then given by the eigenvalues of the stability matrix $\mathcal{M}$.
We also introduce $\nu=-1/\theta_-$ as the scaling exponent for the correlation length defined as (minus) the inverse of the lowest (negative) eigenvalue of the stability matrix.

It is convenient to reformulate the flow equations (\ref{lock}) in terms of the mass coupling in place of the flow of the potential minimum. We define $u'|_{\rho=1}=m^2$. We find
\beq
\begin{array}{lcl}
\partial_t m^2&=&-2m^2+2(m^2)^2+4\lambda m^2\,,\\[1ex]
\partial_t\lambda&=&-\lambda+4\lambda^2+8\lambda m^2\,\\[1ex]
\partial_t\tau&=&12\lambda\tau+12\lambda^2+12m^2\tau\,,
\end{array}
\eeq
and similarly to higher order.  Using the above at the Wilson-Fisher fixed point  \eq{locwffp} we find the exact scaling exponents from the local polynomial expansion as
\begin{equation}\label{localWF}
\theta=-1,1,3,5,7\dots\,.
\end{equation}
This is the well-known large-$N$ result for the scaling exponent $\nu$, given by minus the inverse of the sole negative eigenvalue, meaning $\nu=1$. We observe that the bootstrap hypothesis is confirmed from order to order in the expansion \eq{A}.

For completeness we also compute scaling exponents at the Gaussian and tricritical fixed points. At the Gaussian fixed point we observe that the local flows  display the canonical eigenvalues \beq\label{GaussEV}
\theta=-2,-1,0,1,2,\cdots\,.
\eeq
In particular, this leads to the mean field exponent $\nu=1/2$.
Considering the tricritical fixed points with \eq{loctcfp}
we once more find classical eigenvalues \eq{GaussEV}, independently of the free parameter $\tau$.
The scaling exponents are the same as for the Gaussian which is interesting in so far as these correspond to interacting UV fixed points.\footnote{In models of $4d$ quantum gravity, it has been observed that higher order polynomial invariants of the metric field lead to  an interacting UV fixed point with {\it near}-Gaussian exponents \cite{Falls:2013bv,Falls:2014tra,Falls:2017lst}.}
We also note that the local expansion cannot differentiate between the regions $c>\frac{3\pi}{4}$ and $c<\frac{3\pi}{4}$ as in the case of the Legendre flow. We obtain scaling exponents for all values of positive sextic coupling but require the global solution to rule our the strong coupling region.

\subsection{Singularities and convergence}

Next, we investigate convergence-limiting singularities of the global Wilson-Fisher fixed point  in the complex plane, following  \cite{Litim:2016hlb}.
It is well-known that nearby singularities away from the physical regime $(\rho\ge 0)$  impact on the physical regime and constrain the radius of convergence for local expansions \cite{Morris:1994ki,Falls:2016wsa,Litim:2016hlb,Falls:2016wsa}. It has been observed previously that polynomial approximations of the Polchinski equation at finite $N$ show  notoriously poor convergence \cite{Bervillier:2007tc}.  Here, we  investigate this aspect at infinite $N$. To that end, we evaluate the polynomial expansion \eq{A}, \eq{wfana} to high order. We find that the fixed point equations for all polynomial couplings can be solved iteratively, and exactly, to each and every order. We find that the set of couplings at the Wilson-Fisher fixed point come out with alternating sign. The ratio test can then be applied to estimate the radius of convergence which is found, approximately, to be
\beq\label{Rapprox}
R=1.157\,.
\eeq
To obtain this estimate we have used the first 30 coefficients in the expansion \eq{A} at the fixed point together with extrapolation to infinite order. The two-fold periodicity pattern indicates that the radius of convergence is limited by a singularity (pole or cut) on the negative real $\rho$-axis. This behaviour is qualitatively similar to polynomial expansions of the closely related Wetterich equation where a pair of complex poles dictates the radius of convergence \cite{Litim:2016hlb,Juttner:2017cpr}. The important new result here is the finiteness of the radius \eq{Rapprox}, which is in stark contrast to models with finite $N$, \cite{Bervillier:2007tc}. We conclude that the presence of Goldstone mode fluctuations is responsible for the poor convergence behaviour of polynomial approxiamtions of Polchinski's flow, summarised in \cite{Bervillier:2007tc}.

Some analytical understanding about the singularity can be deduced from the analytical solution \eq{rhowf}. As can be seen in Fig.~\ref{largeNdense},
along the Wilson-Fisher solution (thick white curve) 
we observe that 
$d\rho/du'=1/u''$ vanishes in the regime $\rho<0$, hence corresponding to a singularity in the quartic coupling $u''$. This happens as soon as
\beq\label{trans}
H'(u')=0\,,
\eeq
with $H$ as in \eq{H}. 
In the entire complex plane the transcendental equation \eq{trans} has a unique finite solution located on the real axis where $\rho<0$,
\beq\label{sing}
\left\{
\begin{array}{rcl}
u'_s&=&-1.84638\cdots\\[.5ex]
\rho_s&=&-0.156604\cdots
\end{array}
\right.
\eeq
The nature of the singularity is determined by expanding the exact solution \eq{rhowf} in the vicinity of \eq{trans}, \eq{sing}, leading to
\begin{equation}\label{vicinity}
\rho-\rho_s=\s012 H''_s\cdot (u'-u'_s)^2+{\rm subleading}\,,
\end{equation}
where $H''_s\equiv H''(u'_s)\neq 0$. The linear term in $(u'-u'_s)$ is absent at the singularity due to \eq{trans}. We then find that
$u'(\rho)-u'_s=\sqrt{2(\rho-\rho_s)/H''_s}$
close to the singularity and modulo subleading corrections, meaning that the singularity is of a square-root type
\begin{equation}\label{u2pole}
u''(\rho)=\frac{1}{\sqrt{2H''_s}}\frac{1}{\sqrt{\rho-\rho_s}}\,.
\end{equation}
Since the singularity at \eq{sing} is the unique solution to \eq{trans} for finite fields, and we expect that it dictates the radius of convergence $R$ for expansions of $u'(\rho)$. Specifically, the exact radius of convergence for an expansion of $u'(\rho)$ about the potential minimum, such as \eq{A} at the Wilson-Fisher fixed point, is
\beq\label{R}
R=|\kappa_*-\rho_s|=1.156604\cdots\,.
\eeq
This result is in excellent agreement with the numerical estimate \eq{Rapprox}.

\section{\bf Ultraviolet fixed points}\label{UV}

\label{contsaddlepoint}
In this section, we investigate the set of weakly and strongly interacting ultraviolet fixed points of the theory, their vacuum stability
and the UV conformal window.

\subsection{Asymptotic safety}

The $O(N)$ symmetric scalar theory in $3d$ is superrenormalisable in perturbation theory. At short distances, it is characterised either by a free or by an interacting UV fixed point corresponding to asymptotic freedom or asymptotic safety, respectively. The Gaussian UV fixed point for asymptotic freedom corresponds to the limit where $c\to \infty$ in \eq{solution}, while settings with asymptotic safety correspond to some cases where $c\neq 0$.

Next, we identify the fixed point solutions  \eq{solution} with asymptotic safety.
For general parameter $c\neq 0$, the relation \eq{solution} can be resolved explicitly for $\rho=\rho(u')$ leading to two separate branches. To that end, we introduce the functions
\begin{equation}\label{rhotri}
\rho_\pm(u')=\rho_{\rm WF}(u')
\pm c\, \frac{\sqrt{|u'|}}{(1-u')^{5/2}}
\end{equation}
with $\rho_{\rm WF}(u')$ defined in \eq{rhowf}. For fixed $c$, the fixed point solution is then given by either $\rho=\rho_+(u')$ or $\rho=\rho_-(u')$, depending on the size of $u'$. 
Let us first show that \eq{rhotri} leads to  continuously differentiable fixed point potential $u'(\rho)$. Owing to the definitions, differentiability is automatically guaranteed for all fields except at the point \eq{saddle}.   We 
compute the Taylor expansion in $u'$ around \eq{saddle} to find
\begin{equation}\label{taytcfp}
\rho_\pm=1\pm c\,\sqrt{|u'|}+4u'+\mathcal{O}[(u')^{3/2}]\,.
\end{equation}
Inverting the series we obtain explicit expressions for the function $u'(\rho)$. Let us consider first the case where $u'>0$. We find 
\begin{eqnarray}\label{contp}
\nonumber
u'&=&\frac{1}{c^2}(\rho-1)^2 
+\mathcal{O}[(\rho-1)^3]\,,\\
\label{upp}
u''&=&\frac{2}{c^2}(\rho-1)
+\mathcal{O}[(\rho-1)^2]\,,\\
\nonumber\label{uppp}
u'''&=&\frac{2}{c^2}+\mathcal{O}[\rho-1]\,.
\end{eqnarray}
The results hold true for either sign of $c$, and thereby establishes that $u'$ cannot change sign across \eq{saddle} for any integral curve as soon as $c\neq 0$.  All higher order derivatives continue  to be functions of $c^2$ and so the sign of $c$ does not alter the continuity of polynomial couplings accross $\rho=1$. 
The result also  establishes that the quartic coupling $\lambda$ at $u'=0$ vanishes identically, for any $c\neq 0$. This confirms that the solutions correspond to the line of tricritical fixed point detected earlier. Moreover, the  exactly marginal sextic coupling $\tau$ is given by
\beq\label{tauc}
\tau=\frac{2}{c^2}
\eeq
in terms of the parameter $c$. Next, we consider the large-field asymptotics for either of the branches \eq{rhotri}, also recalling $u'>0$. We find
\beq
\frac{\rho}{c_{\rm crit}\pm c}=\frac{\sqrt{u'}}{(1-u')^{5/2}}+\text{subleading}\,,
\eeq
with $c_{\rm crit}$ given in \eq{cBMB}.  This result states that the asymptotic behaviour is dictated by the sign of $(c_{\rm crit}\pm c)$. Resolving for $u'$ we find
\beq\label{asymptotics}
u'(\rho)=1-\left(\frac{c_{\rm crit}\pm c}{\rho}\right)^{2/5}+\text{subleading}
\eeq
 For  $c_{\rm crit}\pm c>0$, \eq{asymptotics} corresponds to the large-field limit $1/\rho\to 0^+$ whereby $u'\to1$.  This behaviour holds true for all branches $\rho_+(u')$ including in the limit $c=0$, see \eq{asymWF}.  For  $c_{\rm crit}\pm c<0$, \eq{asymptotics} corresponds to the large negative field limit $1/\rho\to 0^-$. The  case  $c_{\rm crit}- c=0$ will be treated separately in Sec.~\ref{BMB}.
Examples for fixed point solutions with $c_{\rm crit}- c>0$ and $c_{\rm crit}- c<0$ are shown in Fig.~\ref{branches} .

\begin{figure*}[t]
	\centering
	\includegraphics[scale=0.44]{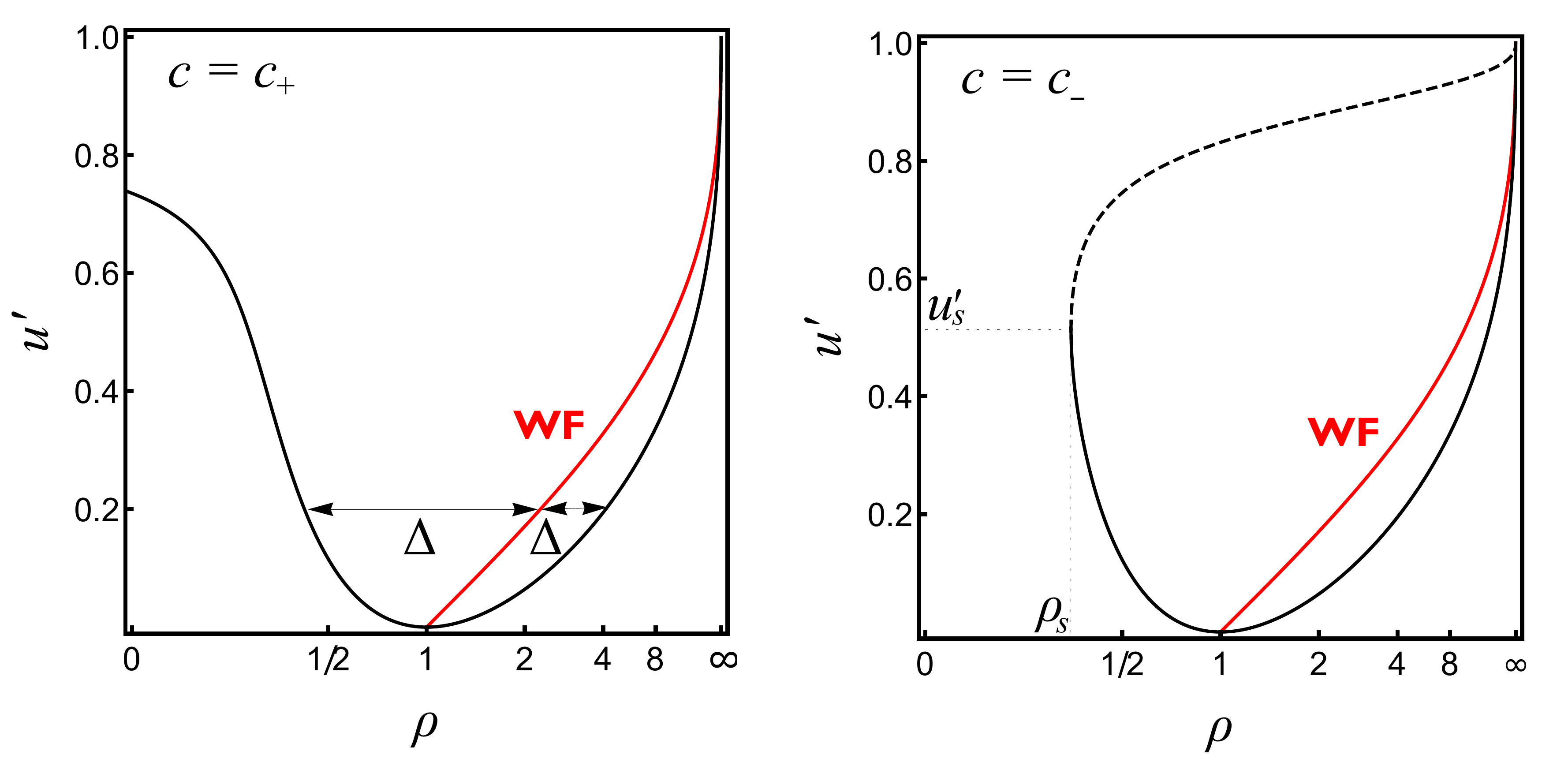}
	\caption{Shown are the Polchinski mass functions $u'(\rho)$ (black lines) for a  weakly coupled UV fixed point   (left panel), and a strongly coupled UV fixed point (right panel) with $c_\pm=c_{\rm crit}\pm 10^{-2}$. 	
	The strong-coupling solution
	exhibits a turning point at $\rho_s$ where $1/u_*''(\rho_s)=0$, and a secondary solution where the fixed point has become multivalued (dashed line).
In either case,
	it is observed that the weak and strongly interacting fixed points follow from the Wilson-Fisher solution (red line) by a constant shift into either direction \eq{rhotri}, with $\Delta=c\,(u
')^{1/2}(1-u')^{-5/2}$.}
	\label{branches}
\end{figure*}
We now turn to the case where $u'<0$. Following the same steps as before, we conclude from \eq{taytcfp}  that the solutions 
\eq{upp}
turn into
\begin{eqnarray}
\label{ncontp}
\nonumber
u'&=&-\frac{1}{c^2}(\rho-1)^2 \\
\label{nupp}
u''&=&-\frac{2}{c^2}(\rho-1)\\
\nonumber
\label{nuppp}
u'''&=&-\frac{2}{c^2}\,.
\end{eqnarray}
modulo subleading corrections in powers of $(\rho-1)$.  We observe that the potential and its derivatives are continuous across \eq{saddle}. Moreover, continuity implies  that $u'$ cannot change sign across the line $u'=0$. 
Evaluating the couplings at $\rho=\kappa_*=1$ we see that once again the quartic coupling vanishes identically. The sextic coupling, this time, becomes negative,
\beq
\tau=-\frac{2}{c^2}\,.
\eeq
Combining this solution with (\ref{uppp}) we see that for $\lambda_*=0$, $\tau_*=\tau$ is a solution for the entire real line consistent with (\ref{loctcfp}). The change in sign of $\tau_*$ between positive and negative $u_*'$ is an important feature of our solution. It means that it is not possible to connect the positive and negative branches of our solution as the sextic coupling contains a discontinuity. Hence we find that the Wilson-Fisher solution $c=0$ is the only fixed point solution for which $u'$ changes sign as a function of the fields.

\subsection{From weak to strong coupling}\label{strongregime}

Tricritical (UV) fixed points are termed ``weakly coupled''  provided the parameter $c>c_{\rm crit}$, corresponding to values for the exactly marginal sextic coupling in the range
\beq
0\le \tau\le \tau_{\rm crit}=\frac{32}{9\pi^2}\,.
\eeq
In the weakly coupled regime, the potential remains finite and well-defined for all fields.  We find that the unique fixed point solution given by
\begin{equation}\label{weak}
\rho=
\left\{
\begin{array}{rcl}
\rho_+(u')\quad\text{if} \quad\rho\ge 1\\[1ex]
\rho_-(u')\quad\text{if} \quad\rho\le 1
\end{array}
\right.
\end{equation}
An example for this is shown in Fig.~\ref{branches}.

\begin{figure*}[t]
	\centering
	\includegraphics[scale=0.44]{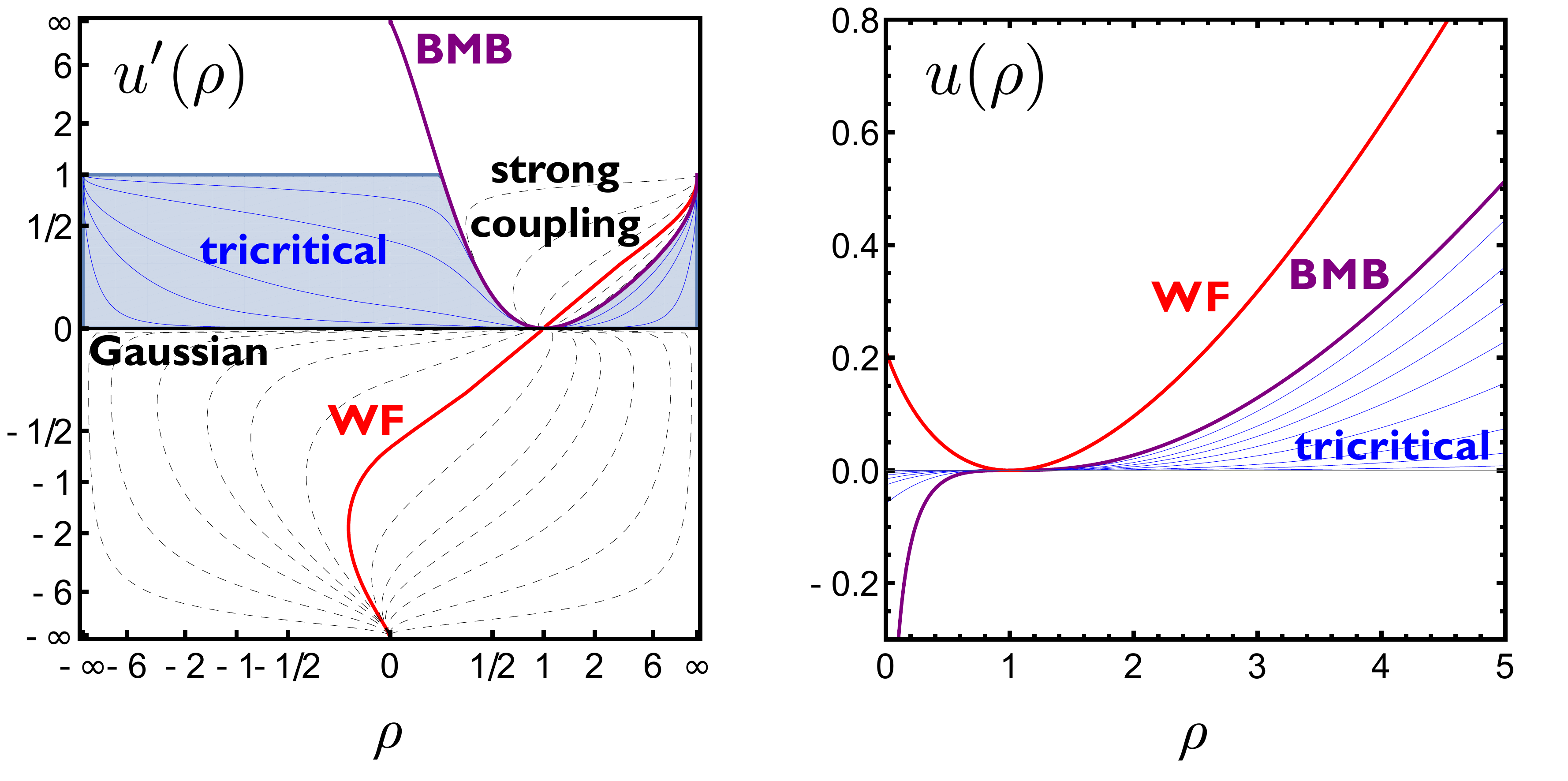}
	\caption{Global fixed point solutions \eq{solution} for the Polchinski mass function $u'_*(\rho)$ (left panel) and the interaction potential $u_*(\rho)$ (right panel) for different values of the free parameter $c$. The red (purple) [black] line corresponds to the Wilson-Fisher (Bardeen-Moshe-Bander) [Gaussian] fixed point. The blue region and blue lines correspond to a family of (tricritical) UV fixed points, and dashed curves denote unphysical multivalued solutions at strong coupling which do not extend over all physical fields.
	The BMB fixed point sticks out, indicating the border between strongly and weakly coupled domains.}
	\label{largeNregions}
\end{figure*}

On the other hand, ``strongly coupled'' fixed points relates to  $c<c_{\rm crit}$, corresponding to values for the exactly marginal sextic coupling in the range
\beq
 \tau_{\rm crit}< \tau\,.
\eeq
The regime of strong coupling is characterised by a turning point in $\rho(u')$ such that $\rho$ becomes multivalued in $u'$ (see Fig.~\ref{branches}, right panel). This turning point occurs when $d\rho/du'=0$ leading to a `Landau-type pole $1/u''(\rho_s)\to 0$ at some point $\rho_s$.

As an aside, we note that singularities and branching points such as this one cannot arise at any finite order in perturbation theory. It is a virtue of the renormalisation group  used here that these features are made visible. Previously, this type of behaviour was observed in supersymmetric $O(N)$ models at finite and infinite $N$, where it leads to cusps for the integrated flow and  the effective potential \cite{Litim:2011bf,Heilmann:2012yf}. It implies a characteristic momentum scale associated to the cusp, similar to the ``Larkin scale''  observed in random field Ising models \cite{Tissier:2011zz}. In ordinary $O(N)$ models, singularities and cusp-like patterns have also been observed   \cite{Litim:2016hlb,Juttner:2017cpr,Marchais:2017jqc}. For further studies of cusps in $O(N)$ models and  the interplay of  finite and infinite $N$ approximations, see   \cite{Litim:2016hlb,Juttner:2017cpr,Marchais:2017jqc,Yabunaka:2017uox,Yabunaka:2018mju}.

Coming back to our setting, we introduce the shorthand notation $u_s'=u'(\rho_s)$ together with (\ref{char2}) to find 
\begin{equation}\label{turning}
\rho_s=\frac{1}{1+4u_s'}\,.
\end{equation}
Inserting this into the branch $\rho_-(u')$ of (\ref{rhotri}) we find that a turning point at $(\rho_s,u'_s)$ arises for theories with $c=c_s$, where
\begin{equation}
c_s=F(u'_s) -4\,\frac{(u'_s)^{1/2}(1-u_s')^{5/2}}{1+4u'_s} \,.
\end{equation}
with $F(u')$ previously defined in (\ref{F}). This expression is monotonous in $u'$ for $0\le u'\le 1$ and reaching its maximum as
\begin{equation}
\lim_{u_s'\rightarrow1}c_s=\frac{3\pi}{4}\,.
\end{equation}
We conclude that the running quartic coupling develops a Landau-type singularity 
for any $0<c<c_{\rm crit}$ in the strong coupling regime. Moreover, the divergence arises in the physical regime $0<\rho_s$. It also entails that  the potential is no longer well-defined for small fields $0\le \rho<\rho_s$.\footnote{This phenomenon has previously been observed using the Wetterich  effective action for the same theory \cite{Litim:2016hlb,Juttner:2017cpr,Marchais:2017jqc},  and for $3d$ supersymmetric $O(N)$ models in the large $N$ limit \cite{Litim:2011bf,Heilmann:2012yf}.} 
Consequently, we find two inequivalent fixed point solutions for every $0<c<c_{\rm crit}$. The first one is given by
\begin{equation}\label{strong1}
\rho=
\left\{
\begin{array}{rcl}
\rho_+(u')&\ \ \text{if}\ \ &\rho\ge 1\\[1ex]
\rho_-(u')&\ \ \text{if}\ \ &\rho_s\le \rho\le 1\quad\text{and}\quad u'\le u'_s\,,
\end{array}
\right.
\end{equation}
while a second solution is given through
\begin{equation}\label{strong2}
\rho=
\rho_-(u')\quad\text{if}\quad\rho_s\le \rho\quad\text{and}\quad u'> u'_s\,.
\end{equation}
In general, if more than one solution are found, the one with the smaller effective potential is the dominant saddle point solution. Notice also  that $u'>0$ for the entire solution \eq{strong2} whereas \eq{strong1} runs through \eq{saddle}. Our results  \eq{weak}, \eq{strong1} and \eq{strong2} for the various independent  fixed point solutions  are summarised in Tab.~\ref{tbranches}.
\begin{table*}[t]
\centering
\begin{tabular}{cccc}
\toprule
\rowcolor{Yellow}
&${}\quad$\bf  weak ${}\quad$  &${}\quad$\bf  critical${}\quad$  &${}\quad$ \bf strong coupling${}\quad$\\ 
\rowcolor{Yellow}
${}\quad$\bf field range${}\quad$&$(c>c_{\rm crit})$&$\ \ (c=c_{\rm crit})\ \ $&$\ \ (0< c<c_{\rm crit})\ \ $\\ \midrule
$1\le \rho\le \infty$&$\rho_+(u')$&$\rho_+(u')$&$\rho_+(u')$\\
\rowcolor{LightGray}
$\rho\le 1$&$\rho_-(u')$&&\\
$0\le \rho<1$&$$&$\rho_-(u')$&$$\\
\rowcolor{LightGray}
$0<\rho_s\le \rho<1$&$$&$$&$\rho_-(u')\ \  (u'<u'_s)$\\
$\rho_s\le \rho\le \infty$&$$&$$&$\rho_-(u')\ \  (u'>u'_s)$\\ 
\midrule
\rowcolor{LightGray}
${}$\quad\# of  independent sols${}\quad$&1&1&2\\
\bottomrule
\end{tabular}
\caption{Domains of validity and branches of the fixed point solutions with $u'\geq 0$. At weak and critical coupling, the junction of $\rho_+$ with $\rho_-$ provides the unique global fixed point solution $\rho(u')$. At strong coupling, two independent global solution exist for all fields $\rho\ge \rho_s$.}\label{tbranches}
\end{table*}
Finally, we note that fixed point solutions with $c\neq 0$ and $u'<0$  display a turning point \eq{turning} at $u'=u'_s$, with  
\begin{equation}
c_s=-\left|F(u_s')-\frac{(1-u_s')^{5/2}}{\sqrt{-u_s'}(1+4u_s')}\right|
\end{equation}
Notice that even the Wilson-Fisher fixed point $(c=0)$ displays a turning point at finite $\rho_s<0$, meaning that the fixed point solution  cannot be defined for arbitrarily large negative fields.

\subsection{Conformal window}\label{conformal}
With these results at hand, we are now in a position to discuss the ``conformal window'' for UV complete quantum field theories (Fig.~\ref{pConformalWindow}). The conformal window is characterised by the sextic coupling $\tau$. Once $\lambda=0$, the coupling $\tau$ is exactly marginal and serves as a fundamentally free parameter of the theory. For negative  $\tau$, the global fixed points have $u'<0$ throughout. In consequence, the vacuum is unstable and a meaningful ground state cannot be identified. For $\tau=0$, the UV fixed point is the well-known Gaussian, corresponding to asymptotic freedom. For $0<\tau<\tau_{\rm crit}$ where the fixed point remains weakly interacting, the theory displays a well-defined {\it interacting} UV fixed point corresponding to asymptotic safety. 
For $\tau=\tau_{\rm crit}$, the UV fixed point additionally displays the spontaneous breaking of scale invariance (Sec.~\ref{BMB}). Within the range
\beq
\tau\in [0,\tau_{\rm crit}]\,,\quad c\in [c_{\rm crit},\infty]
\eeq
the short-distance theory is described by an interacting conformal field theory. At even stronger coupling ($\tau>\tau_{\rm crit}$), the fixed point solutions are incomplete in that they do not extend over the entire  domain of physical fields. These solutions are considered unphysical.

\begin{figure}[t]
	\centering
	\includegraphics[scale=.9]{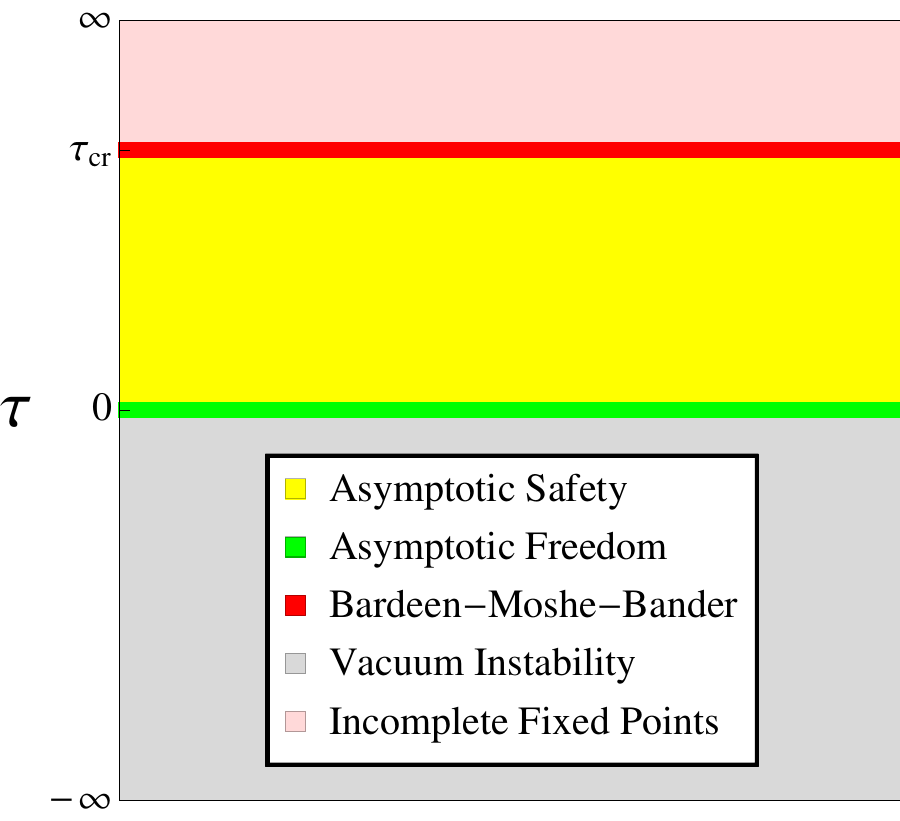}
	\caption{Shown is the exact conformal window with asymptotic safety of large-$N$ scalar field theories, characterised by the bare sextic coupling $\tau$. The regime with asymptotic safety (yellow band) is limited by the Gaussian fixed point ($\tau=0$, green line) and the Bardeen-Moshe-Bander fixed point ($\tau=\tau_{\rm crit}$, red line). Regions with incomplete fixed points (light red band) and unstable vacua (gray band) are also indicated.}
	\label{pConformalWindow}
\end{figure}

\section{\bf Bardeen-Moshe-Bander fixed point}\label{BMB}
We now turn to the  specific value of the parameter $c=c_{\rm crit}$, \eq{cBMB}, where the fixed point \eq{solution} displays the spontaneous breaking of scale invariance, also known as the Bardeen-Moshe-Bander phenomenon \cite{Bardeen:1983rv,Bardeen:1983st,David:1984we,David:1985zz,Matsubara:1987iz,Marchais:2017jqc}. Most notably, the breaking of scale symmetry entails the generation of a light ``dilaton''   or  "scalon" \cite{Gildener:1976ih} in the spectrum.  It has  been speculated that these types of mechanisms may offer a scenario for the origin of the Higgs  in certain extensions of the Standard Model  \cite{Goldberger:2008zz,Bellazzini:2012vz,Cata:2018wzl}.   
The breaking of scale symmetry has also been studied in extensions with supersymmetry or Chern-Simons interactions \cite{Bardeen:1984dx,Eyal:1996da,Litim:2011bf,Heilmann:2012yf,Aharony:2012ns,Bardeen:2014paa}. Here, we explain how this phenomenon comes about within Polchinski's flow equation and in terms of the Polchinski interaction potential \cite{Tetradis:1995br,Litim:1995ex,Comellas:1997tf,Marchais:2017jqc}. We also investigate the Polchinski-Wetterich duality in depth, particularly in view of the breaking of scale symmetry at an RG fixed point.

\subsection{Spontaneous breaking of scale invariance}
By their very definition, fixed points of the renormalisation group imply that dimensionless couplings remain unchanged under a change of RG scale, which entails scale invariance.
Within the RG, and on the level of the dimensionless effective potential $w(z)$ in \eq{both}, the spontaneous breaking of scale invariance can be understood as follows  \cite{Marchais:2017jqc}. In the  limit $k\to 0$, the  physical mass  $m$ of the scalar field in the symmetric phase  relates to the derivative of the quantum effective potential, which itself is related to the fixed point solution as  $W'(z\cdot k)=w_*'(z)\,k^2$,
\begin{equation}\label{mPhys}
m^2\equiv W'(0)= w_*'(0)\,k^2\,.
\end{equation}
Notice that the mass squared $m^2$ is non-zero for any fixed point solution with $w_*'(0)\neq 0$ and $k\neq 0$. This is in accord with the requirements of scale invariance because the mass parameter itself scales proportionally to the RG scale, $m\sim k$. Moreover, as long as $w_*'(z=0)$ at vanishing field remains finite, the mass vanishes in the  limit $k\to 0$.
In our case, this applies for all weakly coupled fixed points where $c> c_{\rm crit}$. However, this conclusion  may be upset provided that $w_*'(0)$ diverges \cite{Marchais:2017jqc}, 
\begin{equation}\label{wprime0}
w'_*(0)\to\infty\,,
\end{equation}
in which  case the physical mass \eq{mPhys} may take any finite value for $k\to 0$, and remains undetermined otherwise.
Specifically, in the physical limit the relation between the classical field $\bar z=z\cdot k$ and the derivative of the effective potential $W'(\bar z)$ for any fixed point solution is given by 
 \beq\label{mass}
 \bar z = \left(c_{\rm crit}
 -c\right) \sqrt{W'}
 \eeq
for any RG scheme \cite{Marchais:2017jqc}, and up to terms subleading in $k^2/W'\ll 1$.  The result establishes that the physical mass at vanishing field 
 must vanish for any fixed point solution with $c\neq c_{\rm crit}$, in accord with exact scale invariance. On the other hand, for $c= c_{\rm crit}$ the mass \eq{mPhys} becomes a free dimensionful parameter which is not determined by the fundamental parameters of the theory, thereby breaking scale invariance. In other words, the mass is discontinuous as a function of $c$, and we conclude that theories with a diverging $w_*'(0)$ at a fixed point lead to the spontaneous breaking of scale invariance \eq{mass} due to the dynamical generation of a mass  \cite{Marchais:2017jqc}. 

\subsection{Polchinski interaction potential}
One might wonder if  the behaviour \eq{wprime0} or \eq{mass} may arise from the Polchinski BMB solution \eq{solution}, and whether the breaking  of scale invariance will again correspond to a small-field divergence on the level of the dimensionless Polchinski mass function,
\begin{equation}\label{uprime0}
u'_*(0)\to\infty\,.
\end{equation}
As has been observed in \eq{asymptotics} (see also Fig.~\ref{largeNdense}), fixed point solutions $u'_*(\rho)$ are bound from above by $u'_*(\rho)\le 1$ for asymptotically large fields $\rho$, provided that $c_{\rm crit}+c>0$. For the Wilson-Fisher fixed point ($c=0$), this bound is related to a pole in field space (Fig.~\ref{pWFcomplex}). Interestingly though, for the BMB case where $c=c_{\rm crit}$, we find that the limit $u'\to 1$ is achieved for finite field $\rho\to\s015$ (see Fig.~\ref{largeNregions}).
Moreover, the fixed point solution has an analytical expansion in $u'$ beyond $u'=1$. We observe that
\beq\label{analytic}
\rho=G(u')\equiv 
\015-\04{35}(u'-1)+\08{105}(u'-1)^2-\0{64}{1155}(u'-1)^3+\dots
\,,
\eeq
where the analytic function $G(x)$ can be expressed in terms of a hypergeometric series 
\beq\label{2F1}
G(x)=\015 - \0{4(x-1)}{35} \, {}_2F_1(1, 3, \s092, 1 - x)\,.
\eeq
It is now evident that the BMB solution can be analytically continued towards $u'>1$, and the full expression for the critical mass function \eq{analytic}, \eq{2F1} then takes the piece-wise closed form
\beq\label{BMBpol}
\rho_{\rm BMB}(u')=\frac{1+\s012 u'}{(1-u')^2}
+\left\{
\begin{array}{lcl}
\displaystyle
-\frac{3}{2}\frac{\sqrt{u'}\,\ln(\sqrt{u'-1}+\sqrt{u'})}{(u'-1)^{5/2}}
&\quad\text{for}\quad{}& \quad \ \ \, 1<u' \quad ({\rm where} \ 0<\rho<\s015)\\[3ex]
\displaystyle
\ \ \frac{3}{2}\frac{\sqrt{u'}(\arcsin\sqrt{u'}-\frac{\pi}{2})
}{(1-u')^{5/2}}
&\quad\text{for}\quad{}& 0\le u'\le 1 \quad ({\rm where} \ \s015<\rho<1)\\[3ex]
\displaystyle
\ \ \frac{3}{2}\frac{\sqrt{u'}(\arcsin\sqrt{u'}+\frac{\pi}{2})
}{(1-u')^{5/2}}
&\quad\text{for}\quad{}&0\le u'\le 1 \quad ({\rm where} \ \ 1<\rho)\,.
\end{array}\right.
\eeq
The expression is analytic except at $\rho=0$. In Fig.~\ref{pPolOpt} (left panel), the Polchinski BMB solution \eq{BMBpol}  is shown for all fields. The three segments of \eq{BMBpol} correspond to the dotted (blue) line, the full (red) line, and the dashed (magenta) line, respectively.
From the explicit result, we deduce that $\rho\to 0$ in the limit $u'\to\infty$, which is the fingerprint for the spontaneous breaking of scale invariance, \eq{uprime0}.
For small fields, the Polchinski mass function $u'$  can be expanded as 
\beq
u'_{\rm BMB}=\sum_{n=-1}^\infty\sum_{m=0}^{n+1}\lambda_{n,m}\,\rho^n(\ln \rho)^m\,.
\eeq
For the first few leading terms, we find
\bea
u'_{\rm BMB}&=&
\frac{1}{2 \rho}
+\0{3}{2} \ln \rho+4-\032\ln 2
 -\0{9}{2} \rho\, (\ln\rho)^2
 - (21-9\ln 2) \rho\ln\rho
\nonumber
\\
\label{BMBsmall}
&& - \left(\0{57}{2}-21\ln 2+\092( \ln2)^2\right) \rho
\eea
The  critical mass function \eq{BMBpol}, \eq{BMBsmall} and the  potential $u_{\rm BMB}(\rho)$ are shown in Fig.~\ref{largeNregions} (magenta lines). They arise as the limit $c\to c_{\rm crit}$ from the finite potentials at the tricritical fixed points. We conclude that  $c= c_{\rm crit}$  is the only case where the Polchinski mass function diverges at vanishing field, meaning that \eq{uprime0} is a  fingerprint for the spontaneous breaking of scale invariance.

Let us now discuss the spontaneous breaking of scale symmetry in terms of the interaction mass. We consider the dimensionful Polchinski mass function $U'(\bar\rho)$ at vanishing field. For general fixed point solution \eq{solution} and  \eq{rhotri} with $c>c_{\rm crit}$, and for $\bar\rho<k$, we have
\beq\label{BMBdim}
\bar
\rho=k\left[
1+k^4\sqrt{U'}\,\frac{F\left({U'}/{k^2}\right)-c}{(k^2-U')^{5/2}}\right]
\eeq
with  $F$ given in \eq{F}. We recall that $0<F(x)<c_{\rm crit}$ for $0<x<1$.   For  $c>c_{\rm crit}$, the second term is negative and $U'(0)$ has a real solution for any $k$. Using \eq{analytic}  it is convenient to isolate the analytic from the non-analytic contributions in  \eq{BMBdim},
\beq\label{BMBdim2}
\bar
\rho=k\left[
G\left(\frac{U'}{k^2}\right)
+k^4\sqrt{U'}\,\frac{c_{\rm crit}-c}{(k^2-U')^{5/2}}\right]
\eeq
where the analytic function $G$ is given in \eq{2F1}. For $c>c_{\rm crit}$ the non-analytic term $\propto (k^2-U')^{-5/2}$ 
 then dictates that $U'$ must remain smaller than $k^2$ for all scales.  It follows that $U'(0)=a\cdot k^2$ with $0<a<1$.
 Hence, the only admissible limit for the mass at vanishing field  and for $k\to 0$ is that of a  vanishing interaction mass, 
$U'\to  0$, as one might have expected for a scale-invariant theory.
On the other hand, at the BMB fixed point \eq{BMBpol} where $c=c_{\rm crit}$, the non-analyticity in the denominator in the second term of \eq{BMBdim} and \eq{BMBdim2} is exactly cancelled by the numerator. In this case,
 the square of the interaction mass $U'$ is no longer constrained to be smaller than the square of the RG scale, and  it follows that
\beq\label{limit}
\bar\rho=\frac{k^3}{2\,U'}
\eeq
in the limit where $k\to 0$, up to terms subleading in $k^2/U'\ll 1$.  We conclude that the interaction mass at vanishing field has become a free dimensionful parameter of the critical theory whereby scale invariance is broken spontaneously. Hence, the mass is discontinuous as a function of the parameter $c$,
and 
the result \eq{BMBdim2} for the interaction potential can be seen as the analogue of the  result \eq{mass} for the effective potential.

\subsection{Polchinski-Wetterich duality}
In order to appreciate similarities and differences with the fixed point  for the dual effective potential \cite{Marchais:2017jqc}, we  compare the main features of the Polchinski interaction potential and the optimised Wetterich effective potential at the BMB fixed point in some detail:

\begin{figure*}[t]
	\centering
	\includegraphics[scale=0.45]{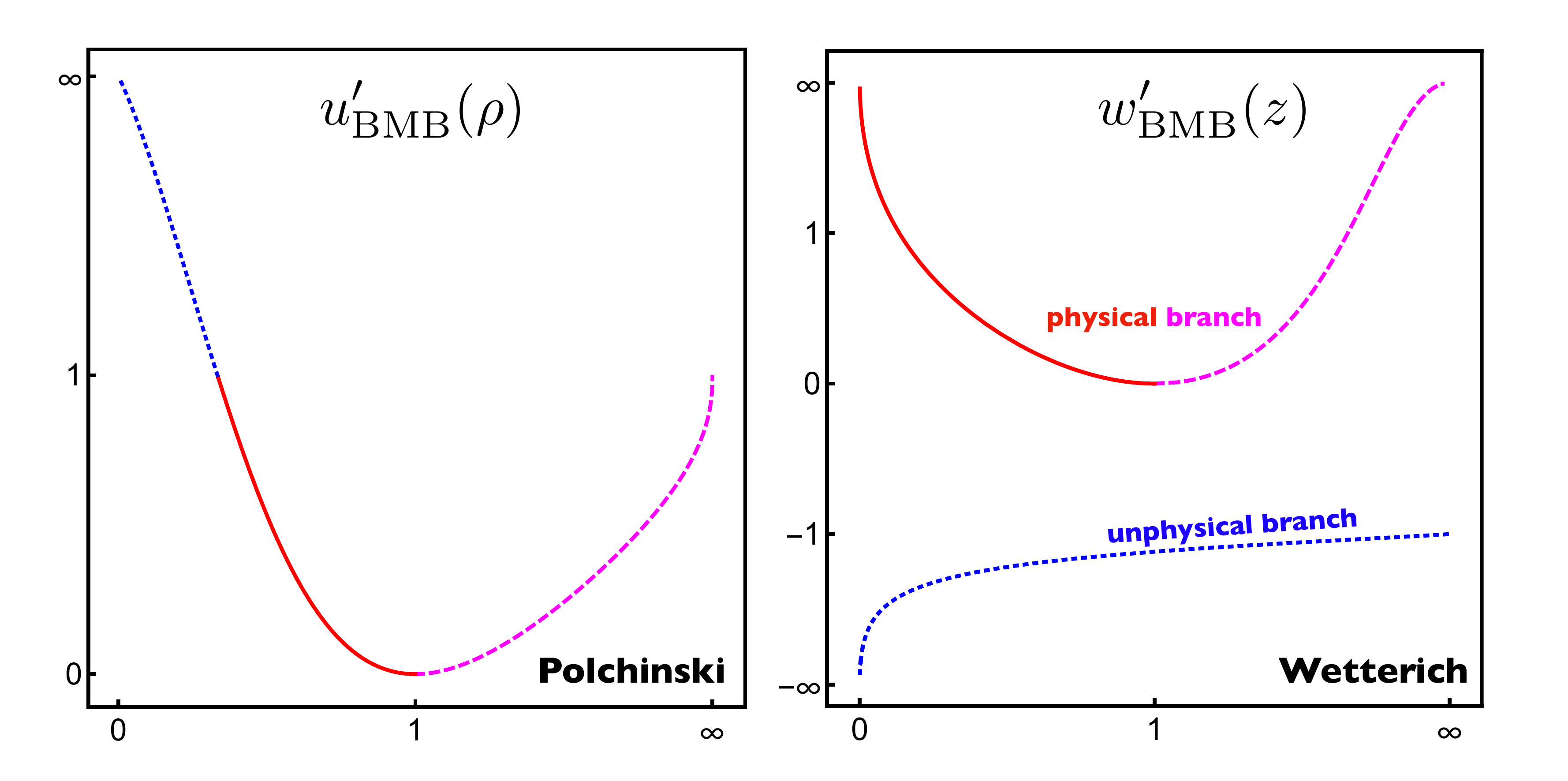}
	\caption{Polchinski-Wetterich duality. Shown are the mass functions at the BMB fixed point for the Polchinski interaction potential (left panel) and for the dual  quantum effective potential  (right panel); axes are rescaled as $x\rightarrow{x}/(1+|x|)$.
	Both mass functions show a non-analytic singularity at vanishing field. The dotted (blue), dashed (magenta) and full (red) segments in both plots are mapped onto each other under the duality  \eq{legpot}. The physical and unphysical branches of the Wetterich flow are connected via analytical continuation \eq{unphysical}. We notice that the divergence in the small-field region of the Polchinski flow originates from the large-field behaviour in the unphysical branch of the Wetterich flow. Conversely, the small field BMB singularity of the Wetterich mass function -- the fingerprint for the spontaneous breaking of scale invariance -- does not correspond to a singularity in the Polchinski mass function.}
	\label{pPolOpt}
\end{figure*}

\begin{itemize}
\item[\it (a)] {\it Unbounded interaction potential.} \\ From \eq{BMBsmall}, owing to the leading term, we conclude that the Polchinski interaction potential is  logarithmically unbounded from below,
\beq\label{BMBln}
u_*\propto  \ln \rho \,.
\eeq
Moreover, since the Polchinski interaction potential is scheme independent to leading order in the derivative expansion,  \eq{dtV}, \eq{vflow}, it follows that the logarithmic unboundedness cannot be removed through a change of RG scheme. This is in marked contrast to the dual effective  potential where the shape of the potential is scheme-dependent. Most importantly, the  effective potential remains bounded for generic regularisation \cite{Marchais:2017jqc}, except for the sharp cutoff limit where it becomes  logarithmically  unbounded  \cite{Bardeen:1983rv,Bardeen:1983st,David:1984we,Marchais:2017jqc}. In \cite{David:1984we}, it has been explained that logarithmic divergences  of the effective potential, such as in \eq{BMBln}, are weak enough to  allow 
for a well-defined ground state.

\item[\it (b)]  {\it Duality.} \\ The Polchinski interaction potential $u_*(\rho)$ from \eq{BMBpol} with \eq{BMBsmall} is dual to the optimised Wetterich  effective action $w_*(z)$ \cite{Litim:2001up,Litim:2002cf,Morris:2005ck} (see also Sec.~\ref{duality}).
In \cite{Marchais:2017jqc}, it has been shown that the  BMB effective potential $w_*(z)$   remains bounded for all fields. In the notation of \eq{duprime}, \eq{both}, for small $z$ we have $w_*'(z)=1/\sqrt{5\,z}\,+$ subleading, and therefore
\beq\label{BMBopt}
w_* \propto \sqrt z\,.
\eeq
It is  noteworthy that one of the potentials is logarithmically unbounded \eq{BMBln} for small fields, while the other  is not \eq{BMBopt}. 
As we shall see, the reason for this is that the duality  does not map the small $z$ region onto the small $\rho$ region. Instead, using the findings from Sec.~\ref{duality}, we observe that the region $w'\to \infty$ and $z\to 0$ from the bounded Legendre effective potential \eq{BMBopt} is mapped onto the vicinity of $u'\approx 1$ and $\rho\approx \s015$ of the Polchinski interaction potential. Due to the analytical continuation accross the point $u'=1$ and $\rho=\s015$, the logarithmically divergent part of the Polchinski interaction potential (where $u'\sim 1/\rho$ for $\rho\to 0$) of \eq{BMBpol} originates from a Legendre effective potential where $w'<0$.
Specifically, the  BMB fixed point solution for the Wetterich effective action \eq{duprime}  is given by  \cite{Marchais:2017jqc}
\beq\label{BMB_opt1}
z_{\rm BMB}(w')=1+\frac{w'}{2(1+w')}+\frac32\sqrt{w'}\arctan\sqrt{w'}+
\left\{
\begin{array}{rl}
-\frac{3\pi}4\sqrt{w'}&\ {\rm for}\ w'>0\ ({\rm where}\ z<1)\\[2ex]
+\frac{3\pi}4\sqrt{w'}&\ {\rm for}\ w'>0\ ({\rm where}\ z>1)
\end{array}
\right.\,.
\eeq
The physical solution \eq{BMB_opt1} has two branches corresponding to large and small fields. We notice that the small-field branch has an analytical continuation from $w'>0$ to negative $w'<-1$. This follows explicitly from its Taylor expansion about $1/w'=0$, which reads
\beq
\label{unphysical}
z_{\rm BMB}(w')=\sum_{n=1}^\infty\frac{n}{2n+3} \frac{1}{(-w')^{n+1}}=\01{5w'^2}\ \,{}_2F_1(2,\s052,\s072,-\s01{w'})
\eeq
in terms of a hypergeometric series.
The analytically-continued branch has $w'<-1$ for all $z\ge 0$. The corresponding potential is globally unbounded for large fields and unphysical.
Hence, the analytically-continued region  $u'\in (1,\infty)$ for physical fields $\rho\in (0,\s015)$  of the Polchinski flow \eq{BMBpol} relates to an unphysical region of the optimised Wetterich flow \eq{unphysical} where $w'\in (-\infty,-1)$ and $z\in (0,\infty)$.  
In Fig.~\ref{pPolOpt} (right panel), the different solution branches  of \eq{BMB_opt1} and \eq{unphysical} are shown by full (red), dashed (magenta) and dotted (blue) lines.

\begin{figure*}[t]
	\centering
	\includegraphics[scale=0.26]{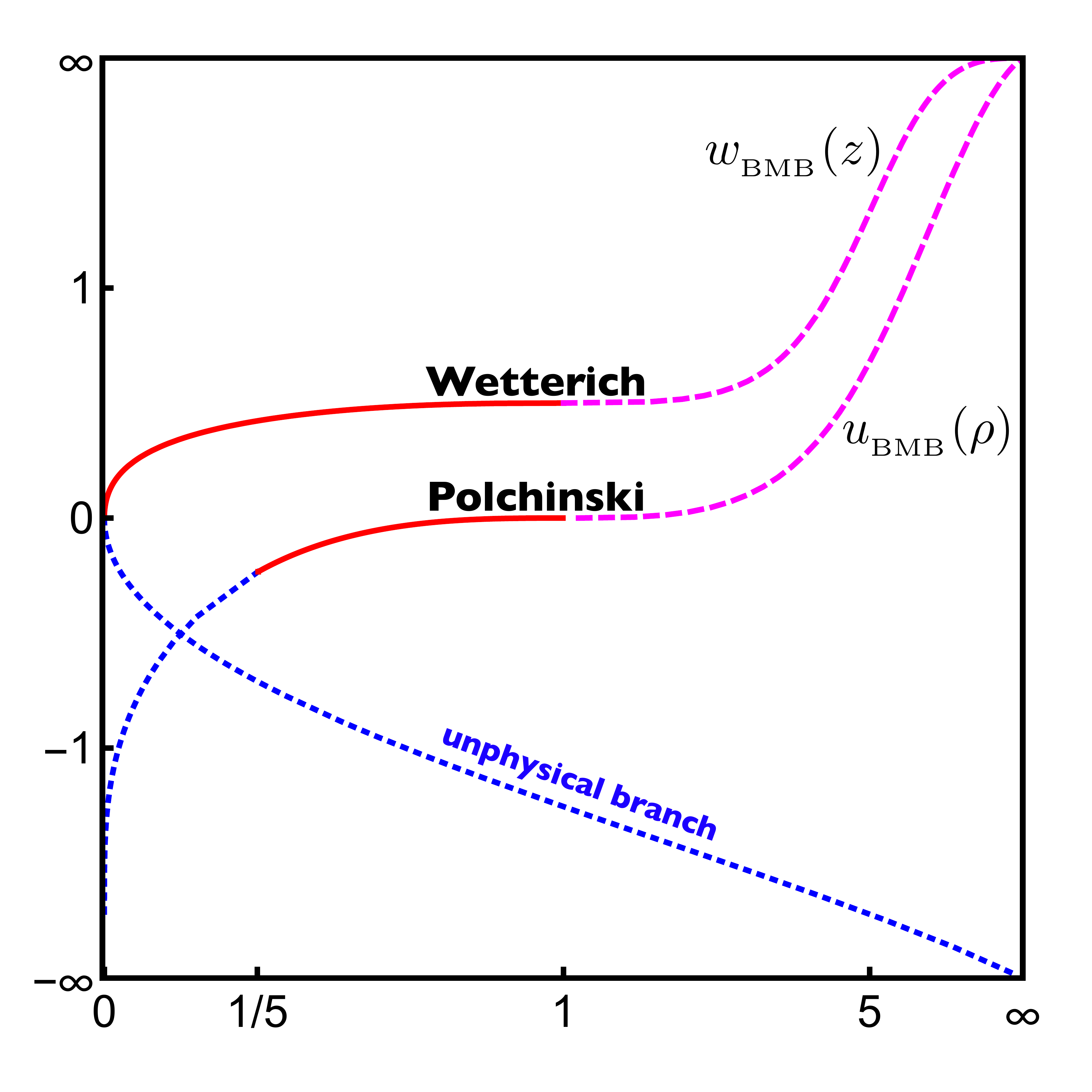}
	\caption{Polchinski-Wetterich duality (cont'd). Shown are the  Polchinski interaction potential $u_{{}_{\rm BMB}}(\rho)$ and the Wetterich  quantum effective potential  $w_{{}_{\rm BMB}}(z)$  at the BMB fixed point; axes are rescaled as in Fig.~\ref{pPolOpt}.
	The dotted (blue), dashed (magenta) and full (red) segments of either potential are mapped onto each other under the duality transformation \eq{legpot}. The small-field region of the Wetterich effective potential, which encodes the breaking of scale invariance,  relates to the mid-field region of the Polchinski interaction potential. The small-field region of the logarithmically unbounded Polchinski  potential  is related to the unphysical (because globally unstable) branch of the Wetterich 
	potential \eq{unphysical}, which is linearly unbounded for large fields.}
	\label{pBMBAll}
\end{figure*}

\item[\it (c)] {\it Polchinski vs Wetterich mass functions.}  \\ Fig.~\ref{pPolOpt} compares the mass function for the Polchinski case $u'(\rho)$  (left panel) and the Wetterich case  $w'(z)$ (right panel) at the BMB fixed point. In the former case the potential is unique \eq{BMBpol}. In the latter case, two inequivalent mass functions arise, a physical \eq{BMB_opt1} and an unphysical one \eq{unphysical}, both of which extend over all fields. Under the duality transform \eq{legpot} the fixed points of \eq{both} and the regions indicated by full (red), dashed (magenta) and dotted (blue) lines in Fig.~\ref{pPolOpt}  are mapped onto each other. Most notably, the 
small-field region of the Polchinski mass function \eq{BMBpol} relates entirely to the unphysical branch \eq{unphysical} of the Wetterich mass function, and the small-field singularity $u'\to1/(2\rho)$ in the former
arises from the large field asymptotics $1+w'\to -1/(2 z)$ of the latter,  \eq{unphysical}. 
On the other hand, the small-field singularity of 
the Wetterich mass function
 \eq{BMBopt} -- the fingerprint for the spontaneous breaking of scale invariance -- does not correspond to a singularity of the critical Polchinski mass function.

\item[\it (d)]  {\it Interaction potential vs effective potential.}\\  Our findings for the different potentials are summarised in  Fig.~\ref{pBMBAll} (colour-coding as in Fig.~\ref{pPolOpt}). We observe that the square-root type small-field non-analyticity, which is responsible for the BMB phenomenon of the effective potential \eq{BMBopt}, relates to finite fields $(\rho=\s015)$ without any divergences within the Polchinski interaction potential \eq{BMBpol}. Conversely, the {small-field} logarithmic unboundedness of the Polchinski interaction potential \eq{BMBln} relates to the {large-field} linear unboundedness ($w\to  -z$ for large $z$) of the unphysical branch of the dual Wetterich effective potential \eq{unphysical}. 

\item[\it (e)] {\it Weak/strong boundary.} \\ In the strong-coupling regime $(c<c_{\rm crit})$, the Polchinski interaction potentials  display a singularity $1/u''\to 0$ at $\rho=\rho_s>0$, \eq{turning}. Also, potentials are not defined in  the field range $0<\rho<\rho_s$. Approaching the weak-strong boundary from above, the BMB-type non-analyticity  of the mass function in field space is located at
\beq
\lim_{c\nearrow c_{\rm crit}}\rho_s=\s015\quad (c\neq c_{\rm crit})\,.
\eeq
For $c= c_{\rm crit}$, however, a finite limit $u''(\rho=\s015)=-{35}/4$ is observed, and the singularity in $u''$, instead, jumps to the origin in field space (Fig.~\ref{largeNregions}). 
A similar pattern is found if the weak/strong boundary is approached from below ($c \searrow  c_{\rm crit}$). Here, the Polchinski mass function, which at vanishing field takes values in the range $(0,1)$, becomes  discontinuous at the point $c =  c_{\rm crit}$ and $1/u'$ jumps from unity to zero  (Fig.~\ref{largeNregions}). 

On the other side, no such discontinuities are observed in the  dual effective potential. At strong coupling, the singularity in $w''$ at $z=z_s$
continuously approaches the origin, $\lim_{c\nearrow c_{\rm crit}} z_s=0$. Similarily at weak coupling, 
the  Wetterich mass function, which at vanishing field covers the entire range $[0,\infty]$, 
 achieves the  limit $1/w'\to 0$ continuously \cite{Marchais:2017jqc}.
Hence, it is precisely for these discontinuities  around the Polchinski BMB fixed point  that the square-root type behaviour of the Wetterich effective potential is mapped onto 
the innocuous-looking point $(u',\rho)=(1,\s015)$ where non-analyticities are no longer visible once $c=c_{\rm crit}$.

\item[\it (f)] 
{\it Broken scale invariance and the origin of mass.} \\ At the BMB fixed point, and irrespective of whether we adopt the Polchinski or the Wetterich version of the flow \cite{Marchais:2017jqc},  the dimensionless critical coupling \eq{cBMB} is turned into a free dimensionful mass parameter, accompanied by the appearance of a light dilaton \cite{Bardeen:1983rv,Bardeen:1983st}. The mechanism whereby the role of a dimensionless parameter is taken over by a dimensionful one, due to fluctuations, is known as dimensional transmutation \cite{Coleman:1973jx}. The BMB phenomenon, arguably, differs from dimensional transmutation in that there is no one-parameter family of RG trajectories representing the various masses of the BMB continuum limit \cite{David:1984we}. Instead, the mass term which breaks scale invariance appears in either case as an additional  free parameter exactly when $c=c_{\rm crit}$, as can be seen from \eq{mass} and \eq{limit} respectively. 

\end{itemize}

To conclude, although both settings \eq{both} are related by a duality-type transformation, we have found that the fingerprint for the spontaneous  breaking of scale invariance is encoded in substantially different manners. Most noticeably, the relevant pattern is given by  \eq{wprime0} and  \eq{mass}, as  observed in the full quantum effective potential. In the Polchinski formulation, however, \eq{uprime0} relates to  an otherwise unphysical branch of the Wetterich flow. Still, it does lead to the spontaneous generation of a mass \eq{BMBdim2}, \eq{limit}. Also the interaction potential remains logarithmically unbounded, irrespective of the scheme, whereas the dual effective potential is bounded. Nevertheless, since both settings describe exactly the same physics, we conclude that the mild unboundedness of  \eq{BMBln} is physically viable, in accord with the result \eq{BMBopt}. Hence, although  the regimes with \eq{wprime0} and \eq{uprime0}  are not directly related under the Polchinski-Wetterich duality \eq{both}, they  both do serve as fingerprints for the  breaking of scale invariance and the spontaneous generation of a mass   \eq{mass}, \eq{limit}.

\section{\bf Universality and phase diagram}\label{PD}
In this section, we discuss universal eigenperturbations and the conformal phase diagram of the theory, including its fixed points and UV-IR connecting trajectories. 

\subsection{Gaussian fixed point}
For scalar field theories in $d>2$ dimensions with finitely many field components and at a Gaussian fixed point, eigenperturbations are known to be generalised Laguerre polynomials \cite{Comellas:1997tf,Litim:2002cf}, which, in some cases, can  be expressed in terms of Hermite polynomials \cite{Hasenfratz:1985dm,Comellas:1997tf,Litim:2002cf}. Specifically, for both versions of the flow  \eq{both} we find the eigenvalue equation
\beq\label{Laguerre}
0=\frac{d+\theta}{d-2}\,f(x)-\left(x-\frac{N}{2}\right)\,f'(x)+x\,f''(x)\,,
\eeq
which is of the  Laguerre-type with eigenvalue $\theta$. Its polynomial solutions are given by the generalised Laguerre polynomials $L_n^\alpha(x)$ with $\alpha= N/2-1$ and integer $n=(d+\theta)/(d-2)\ge 1$ (the case $n=0$ is irrelevant), leading to quantised eigenvalues $\theta$. Eigenperturbations of \eq{both} are  related to $f(x)$ by $\delta u(\rho)=f((d-2)\rho/2)$ for the Polchinski case and $\delta w(z)=f((d-2)z/2)$ for the Wetterich case.  The result \eq{Laguerre} is universal in that it holds true  irrespective of the RG scheme \cite{Litim:2002cf}.

 In the infinite-$N$ limit the analysis is mildly modified.
The eigenvalue spectrum around the free fixed point is identified from the linear perturbations, which at infinite $N$ obey the eigenvalue equation
\beq
\partial_t \,\delta u=-d\,\delta u+[(d-2)\rho-1]\delta u'=\theta\,\delta u\,,
\eeq
where $\delta u$ denotes the eigenperturbation and $\theta$  the eigenvalue. Note also that we have scaled a factor of $N$ into the fields. It follows from straightforward integration that the Gaussian eigenperturbations are given by
\beq\label{monomials}
\delta u\propto e^{\theta t}\,\left(\rho-\01{d-2}\right)^\frac{d+\theta}{d-2}\,.\eeq
Imposing analyticity of eigenperturbations in the fields $\rho$ leads to a quantisation condition for the exponent $(d+\theta)/(d-2)$, which can only take  non-negative integer values. It follows that the quantised eigenvalues are given by 
$\theta_n=(d-2)n-d$  for $n\ge 0$.
The most relevant eigenvalue $\theta_0=-d$ relates to the vacuum energy and is irrelevant in any QFT without gravity. The remaining eigenvalues are simply given by the engineering mass dimension of the highest $\rho$-monomial $\int d^dx \rho^n$  contained 
in the eigenperturbation \eq{monomials}, plus terms of lower powers in the fields due to tadpole corrections  \cite{Altschul:2004yq}.
The result \eq{monomials} also follows from the Laguerre polynomial solutions to \eq{Laguerre} at finite $N$ by observing that the generalised Laguerre polynomials obey
\beq \label{LaguerreN}
\lim_{\alpha\to\infty} \alpha^{-n}L^\alpha_n(\alpha x) =\frac{1}{n!}(x-1)^n\,
\eeq
in the large-$N$ (large $\alpha$) limit. Hence, for the  eigen-perturbations we  conclude that the  Laguerre polynomials are superseeded by shifted field monomials \eq{monomials}, \eq{LaguerreN} in the infinite-$N$ limit. 

\subsection{Interacting fixed points}
\label{scalingglobal}
In order to find the universal scaling exponents  at interacting fixed points, we must find the eigenvalue spectrum of small perturbations. In the infinite-$N$ limit  the global linear eigenperturbations can be calculated exactly without resorting to a polynomial approximation \cite{Comellas:1997tf,Litim:2011bf,Heilmann:2012yf,Marchais:2017jqc}. To that end, we consider  perturbations around the interacting  fixed point $u_*(\rho)$, with
\begin{equation}\label{pertfixed}
u(\rho,t)=u_*'(\rho)+\delta u(\rho,t).
\end{equation}
Inserting \eq{pertfixed} into (\ref{highneqn}) and linearising the equation in $\delta u$ we find
the non-perturbative RG flow for small perturbations
\begin{equation}\label{eigeneqn}
\partial_t\delta u=\left(-d+2\frac{u_*'}{u_*''}(1-u_*')\partial_\rho\right)\delta u\,.
\end{equation}
To obtain \eq{eigeneqn} we have made use of the identity $\rho(d-2+4 u_*')-1=2{u_*'}(1-u_*')/{u_*''}$, which holds true for any non-trivial fixed point solution of  \eq{incneqn} in the large-$N$ limit.
The eigenperturbations obey the eigenvalue equation
$\partial_t\delta u=\theta\delta u$
where $\theta$ denotes the corresponding eigenvalue. Writing $\delta u(\rho,t)=T(t)R(\rho)$ we find 
\begin{equation}
\begin{array}{rcl}
(\ln T)'&=&\theta\\[1ex]
(\ln R)'&=&
\displaystyle
\frac{1}{2}(d+\theta)\left(\ln \frac{u_*'}{1-u'_*}\right)'
\end{array}
\end{equation}
which can be integrated immediately, leading to $T(t)\propto e^{\theta t}$ and $
R(\rho)\propto  [{u_*'}/(1-u_*')]^{\frac{1}{2}(d+\theta)}$. Consequently,
the eigenperturbations take the form
\begin{equation}\label{scaling}
\begin{array}{rcl}
\delta u(\rho,t)&\propto&
\displaystyle
 e^{\theta t} \left(\frac{u_*'}{1-u_*'}\right)^{\frac{1}{2}(d+\theta)}\,.
\end{array}
\end{equation}
 It is interesting to compare our result for the eigen-perturbations at interacting fixed points of the Polchinski  flow with expressions from its dual flow \eq{duprime}  whose exact eigen-perturbations have been given in \cite{Marchais:2017jqc}. In the conventions used here, \eq{both}, they read
\begin{equation}\label{dualscaling}
\delta w(z,t)\propto
 e^{\theta t} \left(w'_*\right)^{\frac{1}{2}(d+\theta)}\,.
\end{equation}
Using the results of Sec.~\ref{duality}, most notably \eq{key}, we conclude that the eigenperturbations of the Polchinski flow $\delta u$ in \eq{scaling} are {\it identical} to the eigenperturbations $\delta w$ in \eq{dualscaling}  of the optimised Wetterich flow, both at free as well as at interacting fixed points. 

Next, we discuss the eigenvalue spectra \eq{scaling} for the present setting where $d=3$.
Once more, we obtain quantised eigenvalues by imposing analyticity conditions, requiring that eigenperturbations are analytical functions of  $\rho-1\ll 1$ in the vicinity of the saddle point (or minimum) $u'(\rho_0=1)=0$.  Recalling \eq{wfana} at the IR fixed point, we conclude that eigenperturbations obey
\beq\label{poly}
\delta u(\rho,t)\propto
 e^{\theta t} (\rho-1)^{\frac{1}{2}(\theta+3)}\,,
 \eeq
modulo subleading corrections of higher order in $(\rho-1)$, and the exponent $\s012(\theta+3)$ must be a non-negative integer. It follows that $\theta$  must take the quantised eigenvalues\begin{equation}\label{EV-WF}
{\rm IR:}\quad\theta=\{-3,-1,1,3,5,\dots\}\,.
\end{equation}
The eigenvalue $\theta=-3$ corresponds to the canonical scaling dimension of the volume term. Without quantised gravity, this eigenvalue is not observable
and cannot appear in the eigenvalue spectrum, and the sole relevant eigenvalue is $\theta=-1$. It leads to the scaling exponent
for the correlation length $\nu=-1/\theta=1$, in accord with known results for the Wilson-Fisher fixed point. The spectrum \eq{EV-WF} agrees with the spectrum derived from the local RG flow \eq{localWF} as it must.

We repeat the analysis for the UV fixed points. The new piece of information is that all UV fixed point solutions display a saddle around $u'_*=0=u''_*$, and behave as $u_*'\propto(\rho-1)^2$ in its vicinity, see \eq{contp}, which  follows from the Taylor expansion \eq{taytcfp} around the saddle point of the potential. Consequently, the eigenperturbations  \eq{scaling} obey \eq{monomials} to leading order in  $(\rho-1)$.
Then, as already discussed in the Gaussian case, in order for eigenperturbations $\delta u$ to be analytic functions of the fields, the exponent $3+\theta$ in \eq{scaling} may take half-integer non-negative values. This leads to the eigenperturbations \eq{scaling} with the discrete set of eigenvalues
\begin{equation}\label{Gauss}
{\rm UV:}\quad\theta=\{-3,-2,-1,0,1,2,\dots\}.
\end{equation}
Once more, in the absence of quantised gravity, the eigenvalue $\theta=-3$ cannot appear in the spectrum. In contrast to \eq{EV-WF}, we observe two relevant and and an  exactly marginal eigenvalue, implying that the UV critical surface is effectively three-dimensional. These are associated to the mass term, the quartic, and the sextic coupling. Moreover, we find the scaling exponent for the correlation length takes the mean-field value $\nu=\012$. The findings \eq{Gauss} agree with the eigenvalue spectrum \eq{GaussEV} detected earlier based on the local RG flows. At $c=c_{\rm crit}$, however, the scaling analysis \eq{Gauss} is no longer applicable due to additional BMB non-analyticities, leading to $\nu=\013$ instead \cite{David:1984we}. For the same reason, hyperscaling relations  such as $d\,\nu=2-\alpha$ involving the scaling exponent $\alpha$ are valid everywhere (that is, for any $c$) except at the BMB fixed point.

We emphasize that the general results \eq{scaling}, \eq{dualscaling} dictate the structure of eigenperturbations \eq{poly} and the   quantisation of eigenvalues. This has two important implications. Firstly,  the number of relevant eigenvalues is necessarily finite. This ensures predictivity, because  the  outgoing trajectories are characterised by only a finite number of free parameters. Secondly,  the structure also ensures that the eigenvalues of invariants with increasing mass dimension are increasingly irrelevant. Thus, these theories are exact examples for the general bootstrap search strategy put forward in  \cite{Falls:2013bv}.

\begin{figure*}[t]
	\centering
	\hskip-1cm
	\vskip-.4cm
	\includegraphics[scale=0.3]{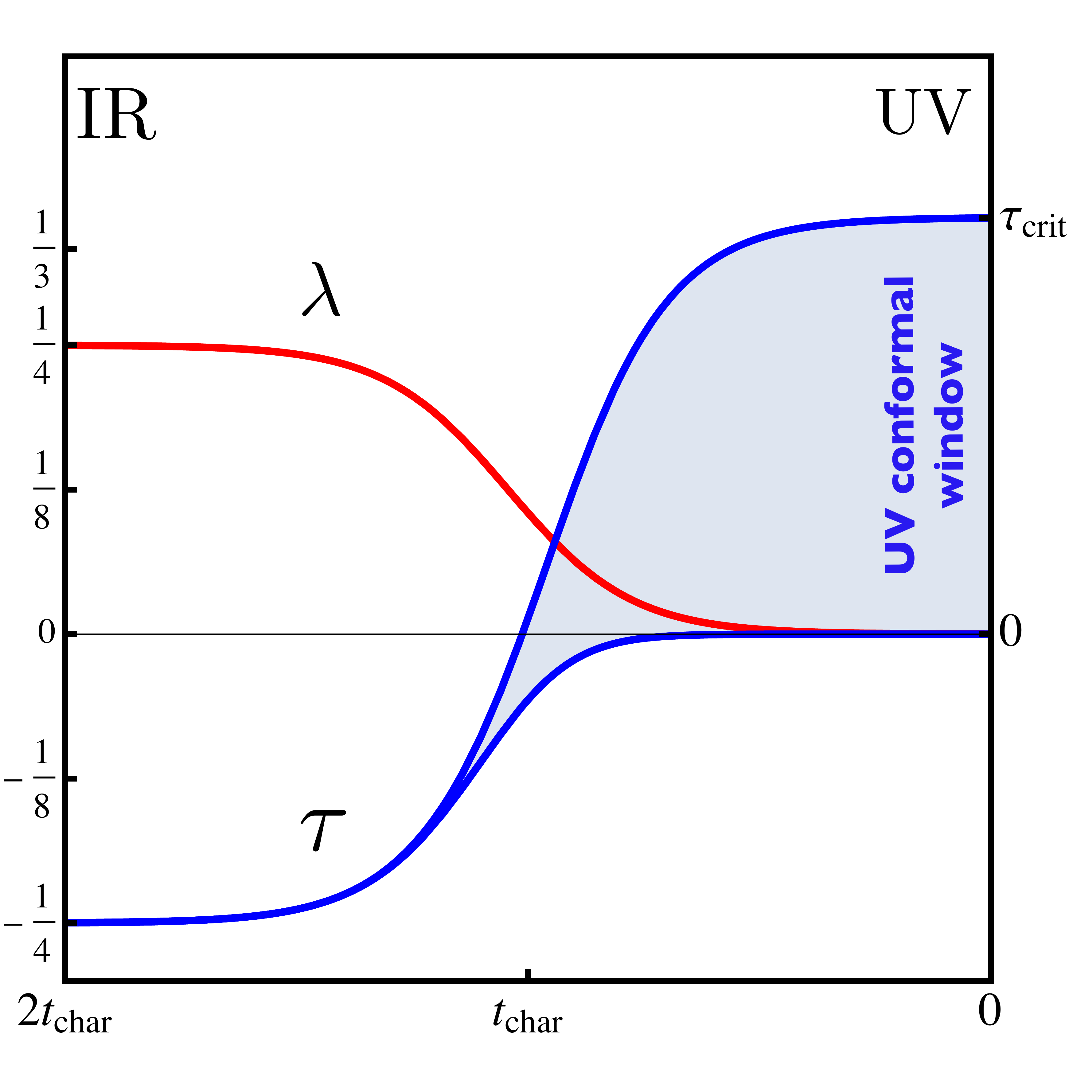}
	\caption{UV-IR crossover from any of the asymptotically safe fixed points to the  unique Wilson-Fisher fixed point. Shown are the running quartic (red) and sextic (blue) couplings along the UV-IR connecting separatrices as functions of $t=\ln (\mu/\Lambda)$ within the  conformal window, with $t_{\rm char}=\ln (\mu_{\rm char}/\Lambda)$  from \eq{muchar}.}
	\label{pChar}
\end{figure*}

\subsection{Asymptotic safety and phase diagram}
The UV critical surface of asymptotically safe theories is effectively three-dimensional (see Sect.~\ref{local} and~\ref{scalingglobal}) and controlled by the mass, the quartic coupling, and the exactly marginal  sextic coupling which serves as a free parameter. Therefore, any UV safe theory can be characterised by the initial deviations $\delta \kappa$ and $\delta \lambda$ from the fixed point couplings at a high scale $\Lambda$,
\beq\label{delta}
\begin{array}{rcl}
\delta\kappa(\Lambda)&=&\kappa(\Lambda)-\kappa_*\\[1ex]
\delta\lambda(\Lambda)&=&\lambda(\Lambda)-\lambda_*
\end{array}
\eeq
and by the value of the sextic coupling $\tau$. All other couplings of the theory including their RG eviolution is now dictated by \eq{delta}, and $\tau$.  Provided that the vacuum expectation value remains at its UV fixed point value $(\delta \kappa=0)$, the phase space of physically viable trajectories is then given by those running out of the UV conformal window towards the Wilson-Fisher fixed point. In the UV, the ground state is in the symmetric phase ($\phi=0$) where it remains for the entire flow provided that $\delta\lambda (\Lambda)>0$.  In this case, it follows from \eq{lambda} that the Wilson-Fisher fixed point acts as an IR attractive ``sink'' for all trajectories. 
The crossover from asymptotically safe UV fixed points to the IR Wilson-Fisher fixed point as a function of \eq{muchar} is shown in Fig.~\ref{pChar} for the entire UV conformal window \eq{window}.  The characteristic  energy scale 
\beq\label{muchar}
t_{\rm char}\equiv \ln (\mu_{\rm char}/\Lambda)=\ln |4\,\delta\lambda(\Lambda)|
\eeq
indicates where the RG flow crosses over from the asymptotically safe UV fixed point towards  either IR scaling at the Wilson Fisher fixed point ($0<\delta \lambda(\Lambda)\ll 1$). It arises through dimensional transmutation from the initial data at the high scale $\Lambda$,  \eq{lambda}. 

\begin{figure*}[t]
	\centering
	\hskip-1cm
	\includegraphics[scale=0.33]{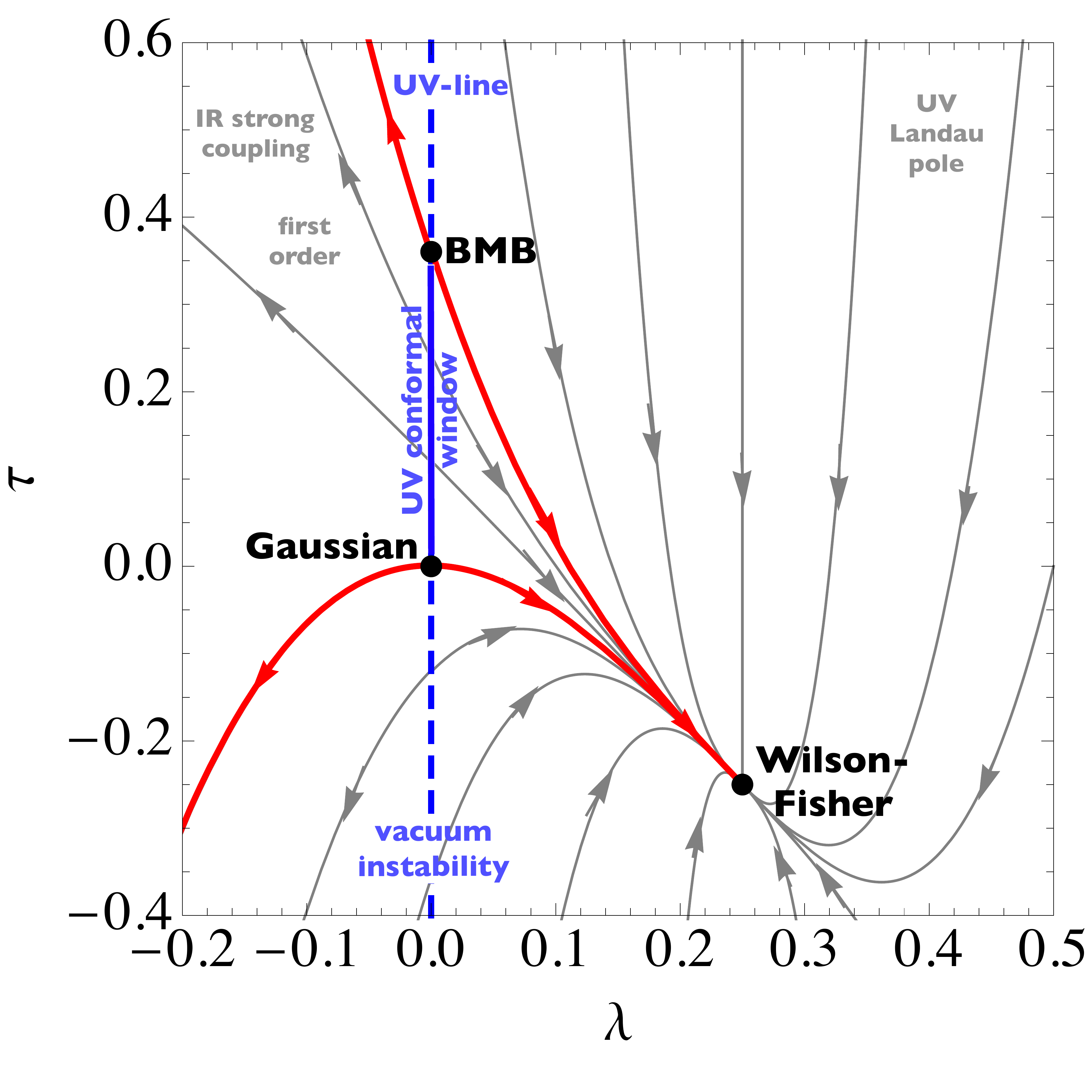}
	\caption{ The phase diagram of $3d$ large-$N$ scalar field theories in the plane of the quartic ($\lambda$) and
		sextic ($\tau$) couplings from the Polchinski  flow, with arrows indicating the flow towards the infrared. The UV conformal window with asymptotic safety $0\le \tau\le\tau_{\rm crit}$ (blue line) is limited by the Gaussian (G) and the Bardeen-Moshe-Bander (BMB) fixed points (full dots). Along the UV-line $(\lambda=0)$, UV fixed points with short-distance vacuum instabilities $(\tau<0)$ or incomplete fixed points $(\tau>\tau_{\rm crit})$ are also indicated  (dashed blue lines). The Wilson-Fisher (WF) fixed point takes the role of an IR attractive sink for asymptotically safe trajectories with $\delta\kappa=0$ and $\delta\lambda>0$.}
	\label{phasediagram}
\end{figure*}

Fig.~\ref{phasediagram} shows a snapshot of the full phase diagram  in the plane of quartic and sextic coupling $(\lambda,\tau)$, with  $\kappa=\kappa_*$. Dots indicate the Gaussian (G), the Bardeen-Moshe-Bander (BMB), and the Wilson-Fisher (WF) fixed points.  The  line $\lambda=0$ relates to potential UV fixed points. The conformal window is  given by the (blue) full  line of asymptotically safe UV fixed points  with
\beq\label{window}
(\lambda,\tau)=(0,0\le \tau\le \tau_{\rm crit})\,.
\eeq
The dashed (blue) lines indicate incomplete fixed points which do not extend over the entire field space ($\tau>\tau_{\rm crit}$), or fixed points with unstable ground states ($\tau<0$). 
Trajectories emanate out of the UV fixed points are shown with arrows pointing towards the IR. 
The red trajectories which connect the Gaussian and the BMB fixed point in the UV with the Wilson-Fisher fixed point in the IR are separatrices, and correspond to the full (blue) trajectories shown in Fig.~\ref{pChar}.

On the other hand, trajectories running towards negative quartic coupling ($0<-\delta\lambda\ll 1$)  will enter a strongly coupled region where couplings approach an  IR Landau pole. 
Along these trajectories, and as a consequence of an increasingly negative quartic coupling $\lambda<0$, the theory may also  undergo a first order phase transition towards a phase with spontaneous symmetry breaking.  Using \eq{lambda}, we find once more that the onset of strong coupling  is characterised by the RG invariant characteristic scale \eq{muchar}.
If  additionally  $\delta \kappa (\Lambda)\neq 0$, trajectories can no longer reach the Wilson-Fisher fixed point in the IR. Instead, trajectories will run towards a low-energy regime in the symmetric phase $(\delta \kappa<0)$ or in a phase with spontaneous symmetry breaking ($\delta \kappa> 0$).
The phase transition towards symmetry breaking may be first or second order, depending on the values of $\delta \lambda$ and $\tau$ at the high scale.
Finally, we note that some of the trajectories terminating at the Wilson-Fisher fixed point do not arise from an UV safe theory. These include trajectories emanating from incomplete fixed points, fixed points with unstable vacua in the UV, or effective models whose UV limit terminates at a UV Landau pole, as indicated in Fig.~\ref{phasediagram}.

\section{\bf Discussion}\label{Discussion}
We have provided a comprehensive study of $O(N)$ symmetric scalar field theories in three dimensions in the limit of infinite $N$ using Polchinski's renormalisation group. The model has been solved analytically both at weak and at strong coupling. Our results offer a detailed understanding of the global running of couplings including all  fixed point solutions, both in the UV and the IR, and for all fields (Fig.~\ref{largeNdense}). Another virtue of the  infinite $N$ limit are insights into the inner working of scalar quantum field theories which are difficult to achieve at finite $N$. These include a complete understanding of convergence-limiting poles in the complex field plane (Fig.~\ref{pWFcomplex}),  interrelations between IR and UV fixed point solutions (Fig.~\ref{branches}), field regions with  Landau type singularities and multivalued solutions (Tab.~\ref{tbranches}), strong coupling effects which constrain the line of asymptotically safe UV fixed points (Fig.~\ref{largeNregions}), and the corresponding UV conformal window (Fig.~\ref{pConformalWindow}). 
Overall, the global phase diagram contains a line of ultraviolet fixed points parametrised by an exactly marginal sextic coupling and bounded by asymptotic freedom at  one end and the Bardeen-Moshe-Bander effect at the other. Separatrices connect the short-distance fixed points with the long-distance Wilson-Fisher one  (Fig.~\ref{pChar}). The phase diagram also shows regions with first order phase transitions and strongly coupled low-energy regimes, and regions without short-distance completion where trajectories  terminate in a Landau pole (Fig.~\ref{phasediagram}). It will be interesting to clarify how these features are modified beyond the $1/N\to 0$ limit.

On the conceptual side, we have  substantiated the duality between Polchinski's and Wetterich's versions of the functional renormalisation group, both on the level of exact solutions (Sec.~\ref{RG}) 
and in view of the  Bardeen-Moshe-Bander  mechanism responsible for the  breaking of scale invariance and the origin of mass (Sec.~\ref{BMB}). In either setting the spontaneous breaking of scale symmetry arises due to the divergence of the respective mass function. 
Unexpectedly though,  
these divergences are  
not mapped onto each other under the Polchinski-Wetterich duality.
Instead, the 
small field region of the Polchinski mass function  relates entirely to an unphysical branch  of the Wetterich mass function whereby the small field singularity  in the former
arises from the large field asymptotics  of the latter  (Fig.~\ref{pPolOpt}). 
In turn, the small field singularity of 
the Wetterich mass function
-- the fingerprint for the spontaneous breaking of scale invariance -- is mapped onto a regular and innocuous mid-field region of Polchinski's mass function.
Another unexpected result is that the Wetterich fixed point potential remains bound from below, while the Polchinski counterpart is logarithmically  unbounded (Fig.~\ref{pBMBAll}). Still, we have confirmed that both sets of results are  equivalent and physically indistinguishable, to the extend that the critical sextic coupling  \eq{tauc} with \eq{cBMB} takes the exact same value  \cite{Marchais:2017jqc}, and that the universal eigenperturbations at all fixed points are identical.

Finally, it is  worth noting  that our $3d$ model with exact asymptotic safety in  scalar field theories 
offers  similarities with asymptotic safety in non-gravitational $4d$ theories. In particular, the  marginality of the sextic coupling implies that asymptotic freedom and asymptotic safety appear as two sides of the same medal, similar to the role played by non-abelian gauge fields in $4d$ theories  \cite{Bond:2018oco}. Strict perturbativity is observed at small sextic coupling, similar to asymptotic safety in certain $4d$ gauge Yukawa theories in a Veneziano limit \cite{Litim:2014uca,Bond:2016dvk,Buyukbese:2017ehm,Bond:2017lnq,Bond:2017suy,Bond:2017sem,Bond:2018oco,Bond:2017tbw,Bond:2017wut}. Also, for large sextic coupling the fixed point becomes non-perturbative and results start resembling large-$N_F$ expansions \cite{Bond:2017wut} and infinite $N_F$ resummations in $4d$ gauge-matter theories \cite{Holdom:2010qs,Litim:2014uca}. It would then be interesting to see whether   solvable $3d$  models may  help to understand UV conformal windows
 of $4d$ gauge-matter theories    at  stronger coupling  \cite{Bond:2017tbw}. This is left for future study.\\[3ex]

\centerline{\bf Acknowledgements}
 Part of this work was performed at the Aspen Center for Physics, which is supported by National Science Foundation grant PHY-1607611. DL   gratefully acknowledges financial support by the Simons Foundation. MJT acknowledges financial support from the CM-CDT under EPSRC (UK) grant number EP/L015110/1.

\bibliography{bib_DFL,bibtex,biblio2_modif2}
\bibliographystyle{JHEP_withtitle}

\end{document}